\documentclass[fleqn,usenatbib]{mnras}

\usepackage{newtxtext,newtxmath}
\usepackage[T1]{fontenc}
\usepackage{ae,aecompl}

\usepackage{graphicx}	% Including figure files
\usepackage{amsmath}	% Advanced maths commands
\usepackage{amssymb}	% Extra maths symbols

\title[]{AGN Feedback and Multi-phase Gas in Giant Elliptical Galaxies}

\author[Wang et al.]{
Chaoran Wang$^1$ \thanks{E-mail: wangcha@umich.edu}, 
Yuan Li$^{1,2}$ \thanks{E-mail: yli@flatironinstitute.org}, 
and Mateusz Ruszkowski$^1$
\thanks{E-mail: mateuszr@umich.edu}
\\
$^{1}$Department of Astronomy, University of Michigan, 1085 S University Ave, 311 West Hall, Ann Arbor, MI 48109\\
$^{2}$Center for Computational Astronomy (CCA), New York, NY, USA\\
}

\date{Accepted XXX. Received YYY; in original form ZZZ}

\pubyear{2018}

\newcommand{\tctff}{t_\mathrm{cool}/t_\mathrm{ff}}
\newcommand{\red}{SPG}
\newcommand{\blue}{MPG}

\begin{document}
\label{firstpage}
\pagerange{\pageref{firstpage}--\pageref{lastpage}}
\maketitle

\begin{abstract}
Recent observations have found extended multi-phase gas in a significant fraction of massive elliptical galaxies. We perform high-resolution three-dimensional hydrodynamical simulations of two idealized elliptical galaxies -- one representing a typical galaxy characterized by initial conditions conducive to the development of thermal instability and the other one less likely to develop thermal instability -- in order to study the development of thermal instability and the formation of multi-phase structures. We analyze the interplay between radiative cooling, momentum-driven AGN feedback, star formation, and stellar feedback from both young and old stars. We find that in one class of elliptical galaxies, the entropy of the hot halo gas rises sharply as a function of radius, and the hot halo is thermally stable and run-away cooling can only happen in the very center of the galaxy. In other class of ellipticals, the hot halo gas has a cooling to free-fall time ratio close to 10, and the non-linear perturbation driven by AGN feedback can cause the hot gas to frequently precipitate into extended multi-phase filaments. Both multi- and single-phase elliptical galaxies experience cooling-driven AGN feedback cycles. Interestingly, AGN feedback maintains the multi- or single-phase nature of the halo but does not turn multi-phase galaxies into single-phase ones or vice versa. Some of the extended cold gas in the multi-phase galaxy also forms young stars. The level of star formation and its spatial distribution are in excellent agreement with {\it Hubble} observations of nearby elliptical galaxies. 
\end{abstract}


%%%%%%%%%%%%%%%%%%%%%%%%%%%%%%%%%%%%%%%%%%%%%%%%%%

%%%%%%%%%%%%%%%%% BODY OF PAPER %%%%%%%%%%%%%%%%%%

\section{Introduction}
The evolution of galaxies is heavily shaped by the supermassive black holes (SMBHs) in their centers \citep[see review by][]{Kormendy2013}. In today's universe, most elliptical galaxies have little ongoing star formation. To maintain this quiescent state, the energy input from the central super-massive black holes (SMBHs) via the active galactic nuclei (AGN) feedback is needed for mainly two reasons. First, the cooling time of the interstellar medium (ISM) in the center of elliptical galaxies is short. 
In the absence of heating, a classical cooling flow is expected to develop in the center of the galaxy \citep{1994ARA&A..32..277F}. The classical cooling flow model over-predicts the star formation rates both in the center of elliptical galaxies and clusters (e.g., \citealt{2001ApJ...557..546D,2005ApJ...635.1031B,2003ARA&A..41..191M,2006ApJ...644..167B,2008MNRAS.385.1186S}). The solution to this ``cooling flow problem'' is some form of heating to offset radiative cooling.
AGN feedback is the most plausible solution due to its high efficiency and self-regulating nature \citep{2007ARA&A..45..117M}.

In addition to solving the ``cooling flow problem,'' AGN feedback is also needed to remove stellar ejecta out of the galaxies. Old stars lose a significant fraction of their original mass to the ISM via stellar winds during the asymptotic giant branch (AGB) phase. The stellar ejecta need to be removed to maintain the observed low density halo \citep{Mathews1971}. Because of the high velocities of the AGB stars themselves (the velocity dispersion of the stars is similar to the velocity dispersion of the hot halo gas), the stellar wind likely thermalizes to the virial temperature of the halo quickly, and this process alone provides a small but non-negligible amount of heating \citep{Mathews1990, 2015ApJ...803...77C} \footnote{Stellar wind heating only increases the total amount of thermal energy per unit volume, but does not increase the specific thermal energy (per unit mass), and thus cannot provide extra energy to do the work required to remove the stellar winds.}. Type Ia SNe (SNIa) explosions inject more thermal energy than thermalized stellar winds, but still do not provide enough energy needed to drive a galactic wind and sweep stellar ejecta out of the galaxy \citep{2003ARA&A..41..191M, 2017ApJ...835...15C}. Cosmological simulations also find that stellar feedback alone cannot prevent late time star formation in massive elliptical galaxies, and that AGN feedback is still needed to maintain the quiescent state \citep{Sijacki2007, TNGSMBH}. 

Accretion onto SMBHs can be dominated by the ``hot mode'' described in the Bondi-Hoyle accretion model \citep{Bondi}, or by the ``cold mode'', assuming that the hot gas first fragments and falls onto the black hole as cold clouds \citep{Balbus89}. Recent Chandra observations of hot X-ray gas in close vicinity of some nearby SMBHs contradict the classical ``Bondi'' model predictions \citep{Baganoff2003, Russell2015}. Theoretical and numerical investigations suggest that hot mode accretion onto SMBHs is inefficient \citep{Yuan2014} and powerful AGN feedback is likely triggered by cold mode accretion \citep{Tremblay2016}. 

There is growing observational evidence for the existence of multi-phase gas in many elliptical galaxies. It has been known for a long time that a significant fraction of elliptical galaxies have dust in their cores \citep{vanDokkum1995, Lauer2005} and the dust is suggested to be linked to AGN activities \citep{Martini2013}. Recent surveys have found both molecular and warm ionized gas in nearby early-type galaxies \citep{Young2011,Pandya2017}. Cold gas has also been detected in significant amounts in the circum-galactic medium (CGM) of early type galaxies \citep{Werk2014}. Some of the cold gas in elliptical galaxies is also turning into stars as is suggested by UV observations \citep{Yi2005}. Recent Hubble observations have directly detected young stars in several nearby elliptical galaxies \citep{2013ApJ...770..137F}. The estimated star formation rate is rather low (on the order of $\sim 10^{-4}~\mathrm{M_\odot \cdot yr^{-1}}$
), but the existence of young stars and star clusters indicate that many elliptical galaxies are not completely ``red and dead.'' 

The effects of different modes of AGN feedback have been studied in many one- and two-dimensional numerical simulations \citep{1997ApJ...487L.105C, 2017ApJ...835...15C, 2018arXiv180301444L}. In the past few years, cosmological simulations have found that mechanical AGN feedback is more effective at ejecting gas out of the halo and suppressing late-time star formation than pure thermal feedback, resulting in early-type galaxy properties that are more consistent with the observations \citep{Choi2012,IllustrisTNG2017}. \citet{2012MNRAS.424..190G, 2017MNRAS.468..751E} have studied the interplay between cooling and AGN feedback in more detail using idealized three-dimensional simulations of elliptical galaxies, but neither work focuses on the multi-phase gas due to limited resolution or treatment of cold gas. 

Recent observations of a sample of nearby elliptical galaxies show that about half of them host spatially extended multi-phase gas, while the rest have single-phase halos \citep{2012MNRAS.425.2731W,Werner2014} \footnote{Note that galaxies with cold gas detected only in the nuclei are classified as single-phase galaxies; only galaxies with extended $\mathrm{H\alpha}$ emission are classified as multi-phase galaxies.}. \citet{2015ApJ...803L..21V} theorizes that in single-phase elliptical galaxies, AGN and SNIa together can drive outflows that sweep stellar ejecta out of galaxies, keeping the ISM in a single phase with only hot gas; in multi-phase elliptical galaxies, SNIa heating is weaker, and thermal instability can develop to form multi-phase gas \citep{McCourt2012, 2012MNRAS.420.3174S}, but precipitation-triggered AGN feedback helps prevent further cooling and keeps the galaxies generally ``quenched.''  

In this paper, we carry out 3D adaptive mesh refinement (AMR) simulations to study AGN feedback and multi-phase gas in two idealized elliptical galaxies based on the observations of NGC 5044 and NGC 4472. NGC 5044 is a multi-phase galaxy (MPG) in \citet{Werner2014}, and NGC 4472 is a single-phase galaxy (SPG). We adopt the momentum-driven mechanical AGN feedback model powered by cold-mode accretion, which has been used in our previous simulations of cool-core galaxy clusters \citep{2014ApJ...789...54L,li2014,Li15} and successfully reproduced many observed features including filamentary multi-phase gas and star formation \citep{Donahue2015, Tremblay2015}. Other important physical processes included in the simulations are self-gravity, radiative cooling, feedback from the old stellar population (stellar wind and SNIa), and star formation and feedback from Type II supernovae. The key questions we try to address are: (1) How do SMBH feeding and feedback relate to thermal instabilities and multi-phase ISM in elliptical galaxies? (2) How does the AGN feedback affect the hot halo gas and drive galactic wind? (3) What causes the difference between multi- and single-phase galaxies? (4) What is the level of star formation and where do young stars form? 

The paper is structured as follows: in Section \ref{sec:methodology}, we describe the simulation setup and how different physical processes are modeled; the main results of the simulations are presented in Section \ref{sec:results}, including the development of thermal instabilities, AGN-driven galactic wind, and star formation; in Section \ref{sec:discussion}, we discuss the caveats of the simulations reflected in the long term evolution of the galaxies, present our resolution and parameter studies, and compare our results with cluster simulations and other elliptical galaxy simulations. We summarize our work in Section \ref{sec:conclusion}.

\section{Methodology}\label{sec:methodology}
Our three-dimensional simulations are performed using the adaptive mesh refinement (AMR) code ENZO \citep{bryan2014} with the ZEUS hydrodynamic method \citep{1992ApJS...80..753S}. 
The simulation domain is a $16~\mathrm{Mpc}^3$ cube with $64^3$ root grids and a maximum of 11 refinement levels. Therefore, the size of the smallest cell is $\Delta x_\mathrm{min}=16~\mathrm{Mpc}/64/2^{11}\approx122~\mathrm{pc}$. We follow the refinement criteria used in galaxy cluster simulations \citep{2012ApJ...747...26L,2014ApJ...789...54L,li2014,Li15}. Detailed descriptions of the refinement criteria can be found in \cite{2012ApJ...747...26L}. Here we only repeat the key points. A non-maximum-refined cell is refined if (1) the cell mass is larger than one-fifth of the gas mass in one cell of the root grid; (2) the ratio of gas cooling time to sound-crossing time of the cell is smaller than 6 (we use a somewhat arbitrary value larger than 1 to better resolve cooling); (3) the size of the cell is larger than four times the local Jeans length. 

In addition to AMR, we place a nested set of static refined boxes in the central area of the simulation domain. The boxes are placed such that the minimum level of refinement increases from 7 at $r=100~\mathrm{kpc}$ to 11 at $r=0.8~\mathrm{kpc}$. This ensures that the AGN jet launching region is always refined to the highest level and that the inner halo is reasonably well resolved even in the absence of cold gas. 

To better resolve the onset of cooling instability and the multi-phase structures, we use the ``super-Lagrangian'' refinement as described in \citet{2012ApJ...747...26L}. This makes the maximum mass of cells decrease with increasing refinement levels. In our standard simulation, the maximum cell mass at $l=11$, our highest level of refinement, is $\sim 1.8\times 10^5 ~\mathrm{M_\odot}$. However many cells in the simulations have much smaller masses ($\sim 10^2~\mathrm{M_\odot}$), partially owning to our static refined boxes.

We model two idealized elliptical galaxies. The initial conditions are described in \S\ref{ss:init}. Important physical processes in our simulations include radiative cooling, self-gravity, momentum-driven AGN feedback, feedback from evolved stars, and star formation, and Type II supernova feedback from young stars. For radiative cooling, we calculate the cooling function derived from the Table 4 of \cite{2009A&A...508..751S} for $T>10^4~\mathrm{K}$ and extend it down to $300~\mathrm{K}$ using the cooling rates given by \cite{1995ApJ...440..634R}. In \S\ref{ss:ph}, we describe in detail how we model the other physical processes.

\subsection{Galaxy Initial Conditions}\label{ss:init}

Our \blue\ and \red\ are modeled to agree with NGC 5044 and NGC 4472, respectively. Both of them are central dominant galaxies of low-mass groups. NGC 5044 has extended multi-phase gas and NGC 4472 is observed to be a single-phase elliptical galaxy. We choose them as representatives of multi- and single-phase galaxies, and refer to them as MPG and SPG throughout the paper instead of using their actual names.

We model the two galaxies in a similar fashion. Both galaxies are initially spherically symmetric and in hydrostatic equilibrium. The gravitational potential for each galaxy consists of self-gravity of the gas and three static components: dark matter, stars, and the central black hole. The dark matter halo is described by an NFW profile \citep{navarro97}, which is characterized by the virial radius $r_\mathrm{vir}$, virial mass $M_\mathrm{vir}$, and the concentration parameter $c$.

The stellar density profile, $\rho_*(r)$ is described by a spherically symmetric de Vaucouleurs profile with a power-law core in the inner region: 
\begin{equation}
\rho_\mathrm{*}(r) =  \left\{\begin{array}{lr}
        \rho_\mathrm{deV}(r; r_e, M_*), & r > r_c \\
        \rho_\mathrm{deV}(r_c; r_e, M_*)(r/r_c)^{-0.9}, & r \leqslant r_c \\ 
        \end{array}\right.,
\end{equation}
where $r_e$ is the effective radius and $M_*$ is the total stellar mass. The approximate de Vaucouleurs profile is adopted from \cite{2005MNRAS.362..197T}. 

The central black hole is treated as a point source of gravity with mass $M_\mathrm{BH}$. We adopt the NFW parameters from \cite{2003ARA&A..41..191M} for the SPG and from \cite{2015MNRAS.448.1979V} for the MPG.
The parameters for modeling the gravity potential for \red\ and \blue\ are listed in Table \ref{tab:grav}. 

\begin{figure}
  \begin{center}      
    \leavevmode
        \includegraphics[width=\columnwidth]{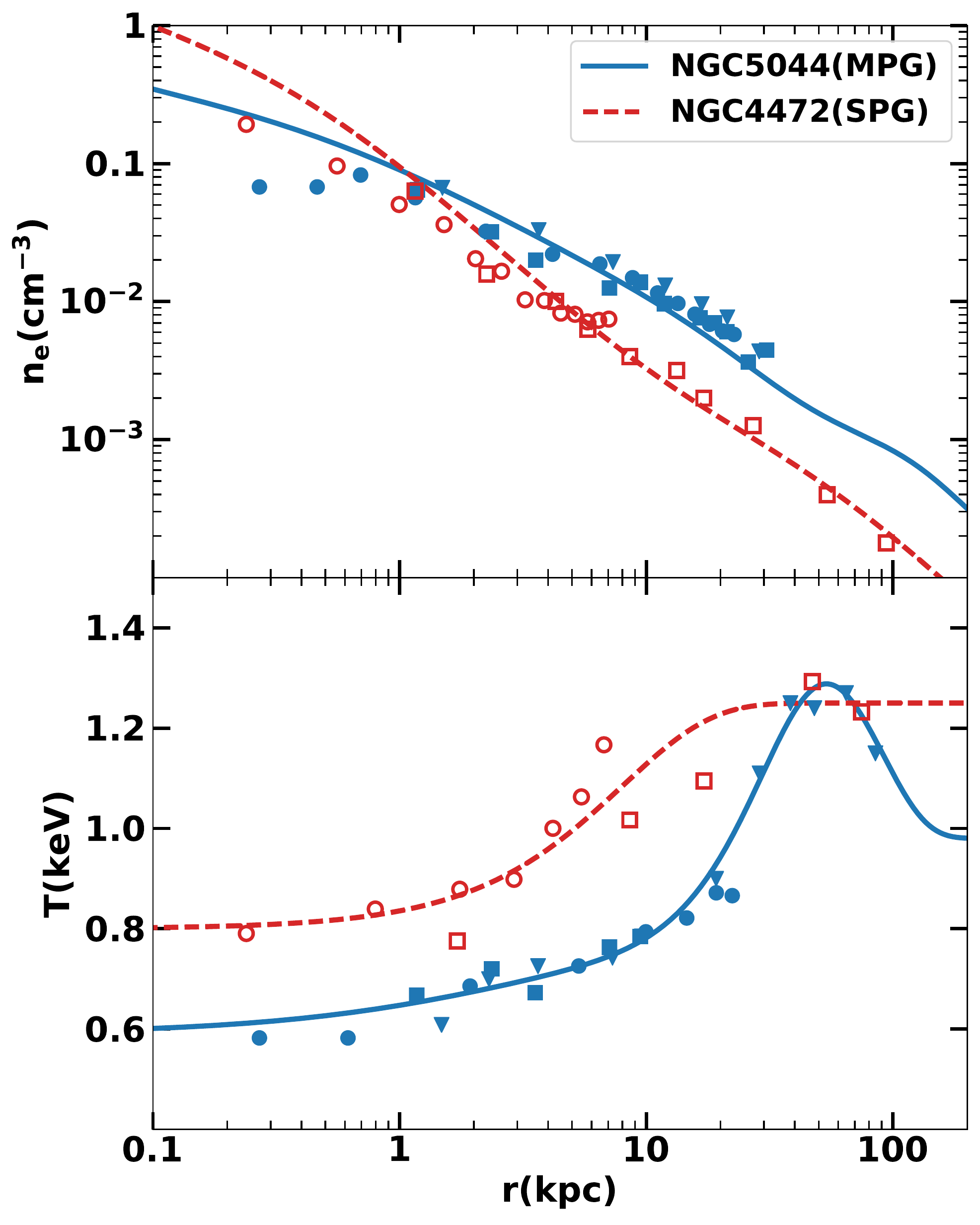} 
       \caption[]{The initial conditions of the two galaxies. Upper panel: the initial electron density profiles of the simulated MPG (NGC 5044, solid blue line) and SPG (NGC4472, dashed red line). The blue points are the observed data for the MPG from \cite{2003ApJ...594..741B} (filled triangles), \cite{2004ApJ...607L..91B} (filled squares), and \cite{Werner2014} (filled circles). The red points are for the SPG from \cite{1996ApJ...471..683I} (open squares) and \cite{2012MNRAS.425.2731W} (open circles). Lower panel: the initial temperature profiles and the observed X-ray data. Color scheme and line styles are the same as in the upper panel.} 
     \label{fig:init}
 \end{center}
\end{figure}

The properties of the hot halo gas are modeled according to the X-ray observations of the two galaxies (\red: \citealt{1996ApJ...471..683I,2012MNRAS.425.2731W}, and \blue: \citealt{2003ApJ...594..741B,2004ApJ...607L..91B,Werner2014}). For each galaxy, we first fit the temperature data using an analytical expression. We then calculate the density profile assuming hydrostatic equilibrium, normalizing it to match the observed data, shown in Figure~\ref{fig:init}. We do not put in any multi-phase gas by hand in the initial setup.

For both galaxies, we set the metallicity of the gas to be a constant solar abundance. This simplistic assumption is consistent with the X-ray observations of the central several tens kpc regions of both galaxies  (for \red, \citealt{2007PhDT.......193A,2006ApJ...646..899H}; for \blue, \citealt{2003ApJ...595..151B,2009PASJ...61S.337K}).

\begin{table*}
\renewcommand{\arraystretch}{1.5}
\label{tab:grav}
\caption{Gravitational potential parameters}
\centering
 \begin{tabular}{ccccccccc} 
%  \hhline{=========}
\hline
 Model & Name & $r_e$(kpc) & $M_*(10^{11}\mathrm{M_\odot})$ & $r_c\mathrm{(kpc)}$ & $c$ & $M_\mathrm{vir}(10^{13}\mathrm{M_\odot})$ & $r_\mathrm{vir}\mathrm{(kpc)}$ & $M_\mathrm{BH}(10^8\mathrm{M_\odot})$ \\
 (1)&(2)&(3)&(4)&(5)&(6)&(7)&(8)&(9)\\
 \hline
 \red & NGC 4472 & 8.57 & $7.26$ & 0.2 & 6.7 &4& 700 & 5.6 \\
 \blue & NGC 5044 & 10 & $3.5$ & 0 & 8.5 & 4 & 900 & 4 \\
 \hline
 \end{tabular}
 \\ \ \\
 \raggedright{{\bf Note.} (1) Model name: \red\ is the single-phase galaxy, and \blue\ is the multi-phase galaxy, (2) Name of the galaxies: \red\ is modeled to agree with NGC 4472 and \blue\ with NGC 5044, (3) Effective radius, (4) Total stellar mass, (5)  Radius of the power-law core of the stellar density profile, (6) Dark matter halo concentration parameter, (7) Total halo mass within the virial radius, (8) Virial radius, (9) Black hole mass. See \S \ref{ss:init} for more details.}
\end{table*}

\subsection{Physical Processes}\label{ss:ph}
\subsubsection{AGN feedback}\label{ss:agnfb}
 The way we model the AGN feedback is similar to that of \cite{2014ApJ...789...54L,li2014}. Here we only repeat the key points.
 \\
 \indent
The SMBH is located in the center of the galaxy. In order to model the black hole accretion, at each time step ($\Delta t$) we calculate the mass accretion rate ($\dot{M}_\mathrm{acc}$) by dividing the total cold gas mass within the central vicinity ($r<500\mathrm{pc}$) by the local characteristic free-fall time (5 Myr). After this time step, $\dot{M}_\mathrm{acc}\Delta t$ of cold gas in this region is removed.
\\
\indent
We model the AGN jets as bipolar outflows launched along the $z-$axis from a pair of jet launching planes. The two jet launching planes are both parallel to the $x$-$y$ plane and located at the center of the simulation domain. Mass ($\Delta m$) is loaded onto the launching planes at each time step following a distribution of $\Delta m\propto\mathrm{exp}\left[\frac{-r^2}{2r_\mathrm{jet}^2}\right]$, where $r\leqslant2r_\mathrm{jet}$ is the distance from the $z$-axis and $r_\mathrm{jet}=1.5\Delta x_\mathrm{min}$. The total amount of launched gas is normalized to be equal to that of the gas removed in the accretion process, namely $\int_{r\leqslant 2r_\mathrm{jet}} \Delta m = \dot{M}_\mathrm{acc}\Delta t$. That is, we assume that only a very small amount of gas is actually accelerated onto the SMBH.
\\
\indent
The jet power is given by
\begin{equation}
P_\mathrm{jet}=\epsilon \dot{M}_\mathrm{acc}c^2.
\label{eq:pjet}
\end{equation}
The feedback efficiency $\epsilon$ is 0.5\% in our standard runs. We further assume that 10\% of the jet power is thermalized (put in as thermal energy) and the remaining 90\% is put in as kinetic energy. 
In our standard simulations, the jet precesses at a small angle ($\theta=0.15$ radian) with a period of 10 Myr. This small precession angle and the exact value of the thermalization fraction of the jet do not have a significant impact on the simulation results. We discuss the impact of the choice of parameters, including the feedback efficiency, in Section~\ref{sec:discussion2}.

\subsubsection{Feedback from old stars}\label{ss:sfb}
For the feedback from old stellar population, we consider the mass and energy input from stellar wind and energy feedback from type Ia supernovae (SNe Ia). The mass loss from old stars are described by a specific mass ejection rate $\alpha=10^{-19}~\mathrm{s^{-1}}$, so that the amount of the ejected matter per unit volume per unit time is $\alpha\rho_*$. The ejected matter is assumed to be thermalized to the stellar virial temperature \citep{2003ARA&A..41..191M}. 
We adopt $\sigma=300~\mathrm{km\cdot s^{-1}}$ which is a typical value for group centrals. 

The SNe Ia feedback from old stars is modeled by injecting pure thermal energy to the ISM. The energy injection is azimuthally uniform, with its rate proportional to the stellar density. The energy input is $10^{51}$ ergs per SN explosion. The specific SNe Ia rate is $3\times10^{-11}~\mathrm{kyr^{-1}\cdot \mathrm{M_\odot}^{-1}}$, which is the value used in \citealt{2015ApJ...803L..21V} and is broadly consistent with the observations \citep{2012MNRAS.426.3282M}. With these parameters, SNIa heating rate is about 5 times the heating rate of the thermalized stellar wind.

\subsubsection{Star formation and stellar feedback}\label{ss:sf}
Star formation and feedback from young stars are modeled in the same way as in \citet{Li15}. We model star formation following \cite{CenOstriker92} with stochastic formation of star particles. A star particle is created if the following criteria are satisfied in a cell: (1) the gas is denser than a critical density (we use $1.67\times10^{-24}~\mathrm{g\cdot cm^{-3}}$ in our simulations), (2) the cell mass exceeds the local Jeans mass, (3) the flow is convergent, and (4) $t_\mathrm{cool}/t_\mathrm{collapse}<1$ where $t_\mathrm{collapse}=\sqrt{3\pi/(32G\rho_\mathrm{gas})}$. The mass of the created star particle is $m_*=0.02m_\mathrm{cell}\Delta t/t_\mathrm{collapse}$ if $m_*>m_\mathrm{*min}=10^{5}~\mathrm{M_\odot}$, where $m_\mathrm{cell}$ is the cell mass and $\Delta t$ is one computational time step. For cells which meet the four criteria but have $m_*<m_\mathrm{*min}$, star particles are formed stochastically with a possibility of $m_*/m_\mathrm{min}$; and the mass of the particle is $\mathrm{min}(0.8m_\mathrm{cell}, m_\mathrm{*min})$. 
This stochastic procedure is used to prevent forming too many star particles. 
We further consider stellar feedback including mass loss and Type II supernova from the created stars. 
For the mass loss, we let the created stars return 25\% of their mass to the gas phase; for Type II supernova feedback, $10^{-5}$ of their rest-mass energy is injected as thermal energy to local cells.

\section{Results}\label{sec:results}

In this section we discuss the properties of the two simulated galaxies over 1.5 Gyr. \S \ref{ss:result1} describes the general gas evolution in \blue\ and \red, and in \S \ref{ss:result2} we analyze the development of thermal instabilities and the formation of cold gas in the two system. In \S \ref{ss:result3} we discuss the how gas sweeping is achieved in the two simulated galaxies and compare it with the theoretical analysis in \cite{2015ApJ...803L..21V}.  We discuss star formation in our simulated galaxies and compare it with observations in Section~\ref{ss:result4}, and in Section~\ref{ss:result5} we discuss the velocity dispersion of the hot ISM.  

\begin{figure*}
  \begin{center}
    \leavevmode
    \includegraphics[width=0.7\textwidth]{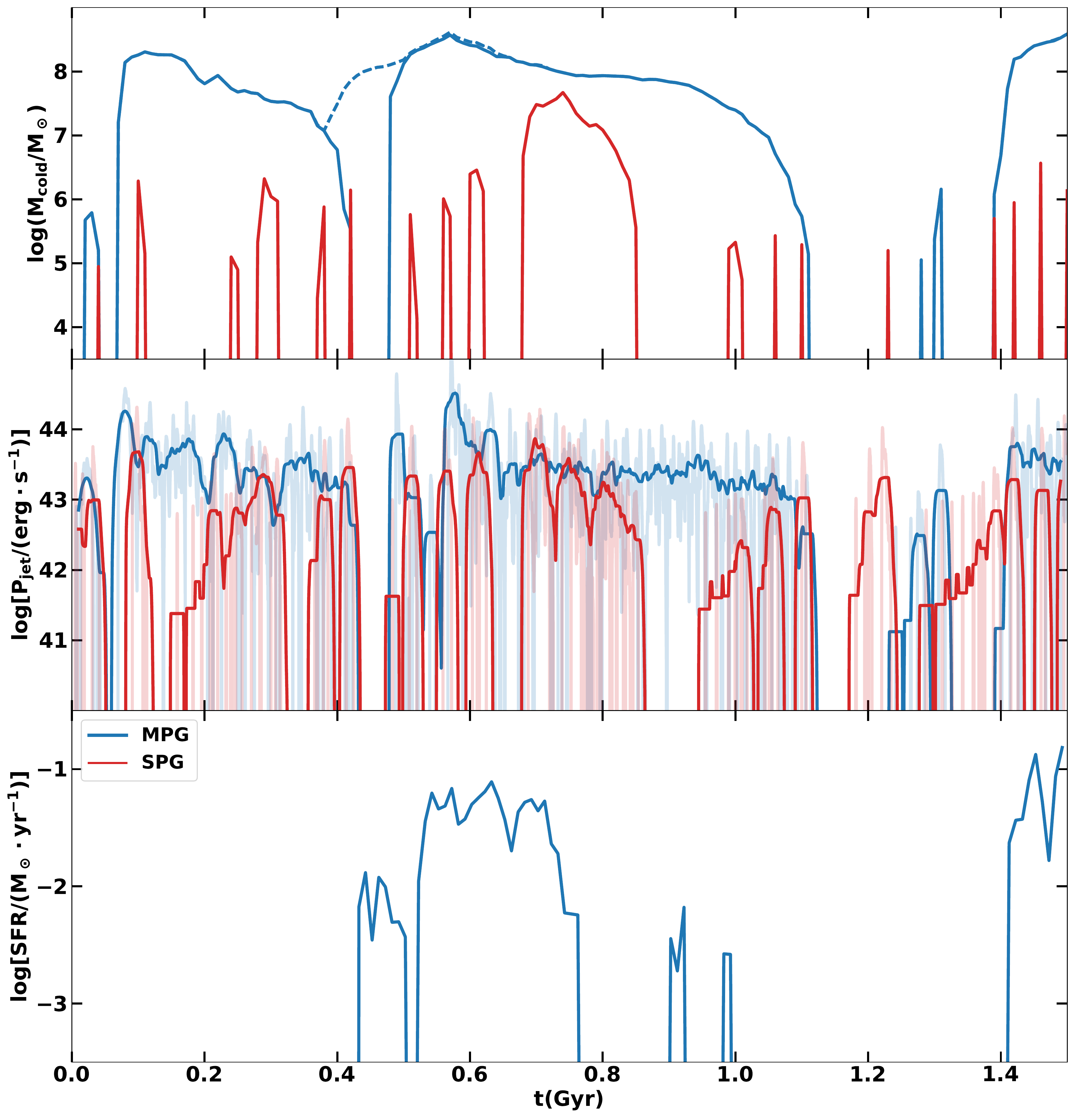}  
       \caption[]{From the top to bottom panel: the evolution of cold gas mass, jet power, and star formation rate in the two simulated galaxies. Red color corresponds to the \red\ and blue to the \blue. In the top panel, the solid blue line corresponds to the cold gas within $r<10$ kpc region; and the dashed line corresponds to the total amount of the cold gas in the galaxy. Simulation data are sampled every 10 Myr. For clarity, we show the instantaneous jet power in lighter color in the middle panel, and add solid lines that are averaged with a 20-Myr moving window. 
       }
     \label{fig:time_evo}
  \end{center}
\end{figure*}

\subsection{General properties of the evolution of the galaxies}\label{ss:result1}
\subsubsection{Multi-phase Galaxy (NGC 5044)}\label{sec:result_MPG}
\begin{figure*}
  \begin{center}
    \leavevmode
        \includegraphics[width=.7\textwidth]{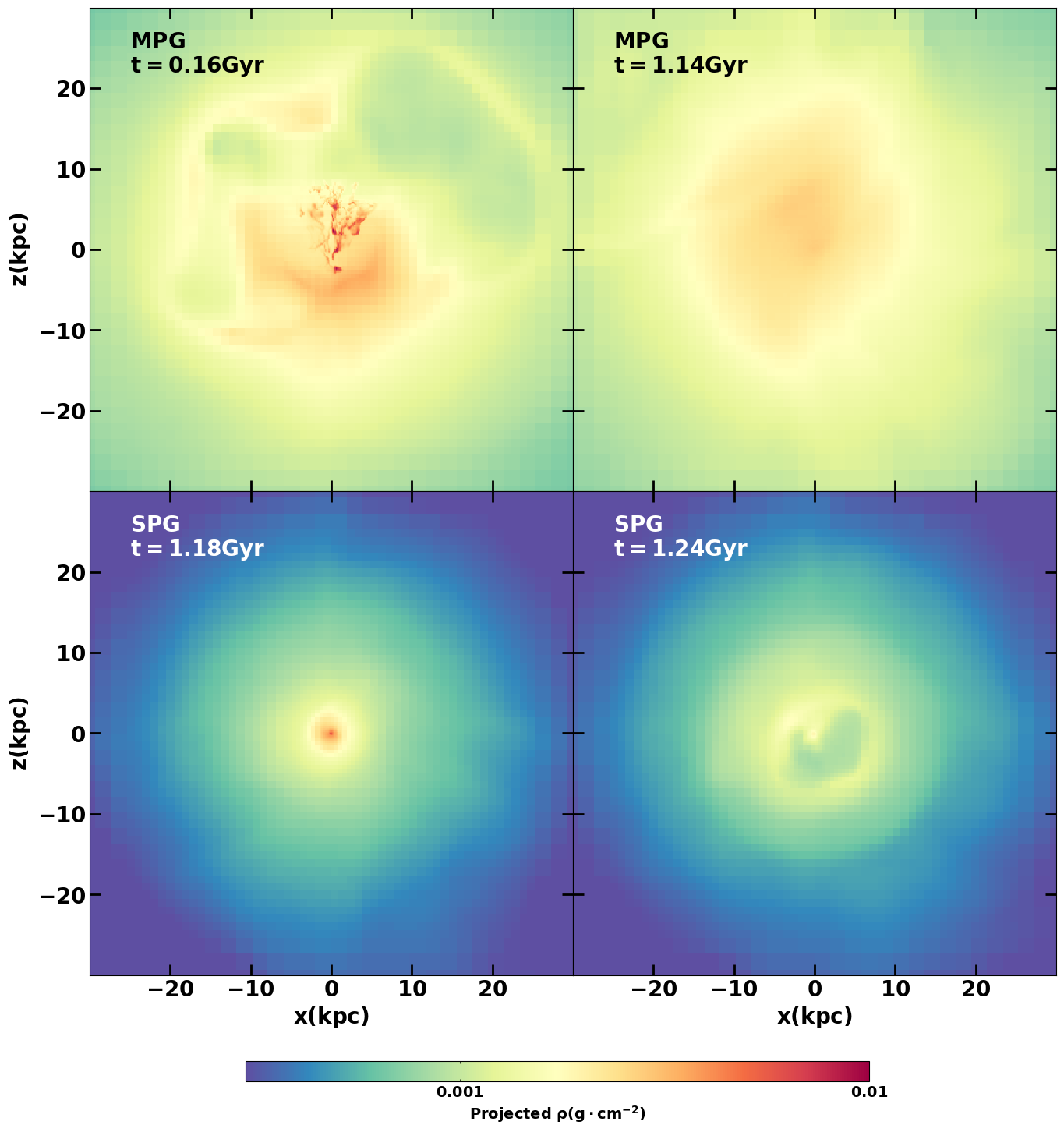} 
       \caption[]{
       Snapshots of projected gas density in the central $(60\mathrm{kpc})^3$ regions of \blue\ (top panels) and \red\ (bottom panels). The projection is along the $y-$axis and the AGN jets are along the $z-$axis.
Animations can be found at \url{https://vimeo.com/266890265} (\blue) and \url{https://vimeo.com/266890473} (\red).
       }
     \label{fig:sx}
  \end{center}
\end{figure*}

In the \blue, the system goes through several major precipitation cycles with AGN activities and extended multi-phase gas. At the beginning of the simulation, AGN feedback is off. Radiative cooling causes gas to condense in the center of the galaxy almost immediately after the simulation starts. This is because in our initial setup, cooling time decreases towards the center monotonically. 
If we keep AGN feedback off, a classical cooling flow develops in the center of the galaxy, which we have verified numerically. In our standard run with AGN feedback, condensation near the SMBH triggers AGN feedback, which quickly heats up the gas in close vicinity to the SMBH. This small amount of energy injection, however, only delays further condensation by $\sim 30~\mathrm{Myr}$ as is shown in Figure~\ref{fig:time_evo}. At $t\approx70~\mathrm{Myr}$, a major precipitation event occurs in the center of the galaxy with a multi-phase structure that extends out to $\sim$10 kpc, which triggers a strong AGN outburst. The gas falls toward the SMBH, and swirls around it as it gradually feeds the SMBH. Meanwhile, more gas continues to precipitate within $r < 10~\mathrm{kpc}$ and continues to power AGN feedback. The mechanical AGN feedback generates shock waves that heat up the ISM as they propagate through the halo. These shock waves also create low density, high temperature regions that resemble X-ray bubbles often observed in these systems (see a snapshot of the projected gas density during this period in the first panel of Figure~\ref{fig:sx}). Interestingly, even though star formation is allowed to happen, no star particle forms in this first cycle of precipitation in spite of the presence of multi-phase gas (third panel of Figure~\ref{fig:time_evo}). 

At $t\approx 400~\mathrm{Myr}$, all the cold gas near the SMBH has been accreted or turned into stars, which turns off AGN feedback. At the same time, a parcel of low entropy gas that has been lifted up by previous AGN activities starts to condense at $r\approx 30~\mathrm{kpc}$. The cold gas falls to the center of the galaxy, forming stars on its way, and reignites AGN feedback, which marks the beginning of the second cycle. Precipitation continues until all the cold gas settles to a clumpy rotating disk around the SMBH at about $t\approx800~\mathrm{Myr}$. The disk then shrinks as star formation and accretion onto the SMBH continues, and eventually vanishes at $t\approx1.1~\mathrm{Gyr}$, which shuts off AGN feedback again. This is the end of the second cycle. The top right panel of Figure~\ref{fig:sx} shows this brief quiescent phase. 

The third cycle begins at $t\approx 1.2~\mathrm{Gyr}$ in ways very similar to the first cycle: a small amount of condensation first happens in the center, triggering a brief AGN outflow that is followed by a major precipitation and AGN outburst. We focus on the first $1.5~\mathrm{Gyr}$ of the simulation and leave the evolution of the galaxy after this point for discussion in Section~\ref{sec:discussion1}. 

The cold gas mass peaks at few times $10^8~\mathrm{M_\odot}$, and the jet power mostly varies within $10^{43}\sim10^{44}~\mathrm{erg\cdot s^{-1}}$ (Figure~\ref{fig:time_evo}). Note that the general cyclical behavior is not sensitive to simulation parameters, but details are. For example, a very small change in one simulation parameter can result in a noticeable difference (a factor of two) in the exact duration of each cycle, the exact amount and spatial extension of the multi-phase gas, and whether precipitation first starts at the center or off center (though the first condensation in the simulation always starts in the center). 

\subsubsection{Single-phase Galaxy (NGC 4472)}\label{sec:result_SPG}
The SPG evolves quite differently from the MPG. As in the case of the MPG, runaway cooling also happens first in the center of the galaxy, and triggers AGN feedback, but this does not lead to a major precipitation event with the formation of extended multi-phase gas. Instead, after the cold gas is quickly accreted onto the SMBH and AGN is turned off, condensation occurs again in the center of the galaxy, with no spatially extended multi-phase gas present. This central condensation ignites the AGN again and the galaxy continues to go through cycles of central cooling and AGN feedback. The bottom panels of Figure~\ref{fig:sx} show the projected gas density of the SPG when the cold gas is about to turn on the AGN (bottom left panel) and when cold gas is gone right after an AGN outburst (bottom right penal).
The peak of the AGN power approaches $10^{44}~\mathrm{erg\cdot s^{-1}}$, only slightly lower than the MPG, but the duration of each cycle is much shorter, typically lasting only tens of Myr or even shorter, followed by several tens of Myr pause (Figure~\ref{fig:time_evo}). 

Within 1.5 Gyr, the SPG only shows extended multi-phase gas at $t\sim 700~\mathrm{Myr}$. The multi-phase structure reaches its maximum spatial extent of $\sim 12$ kpc, and the amount of cold gas rises to $6\times 10^7~\mathrm{M_\odot}$. Except for this brief moment of spatially extended cooling, the galaxy never hosts more than $10^7~\mathrm{M_\odot}$ cold gas, and the cold gas is always only found in the nucleus of the galaxy.

\begin{figure*}
  \begin{center}
    \leavevmode
        \includegraphics[width=.7\textwidth]{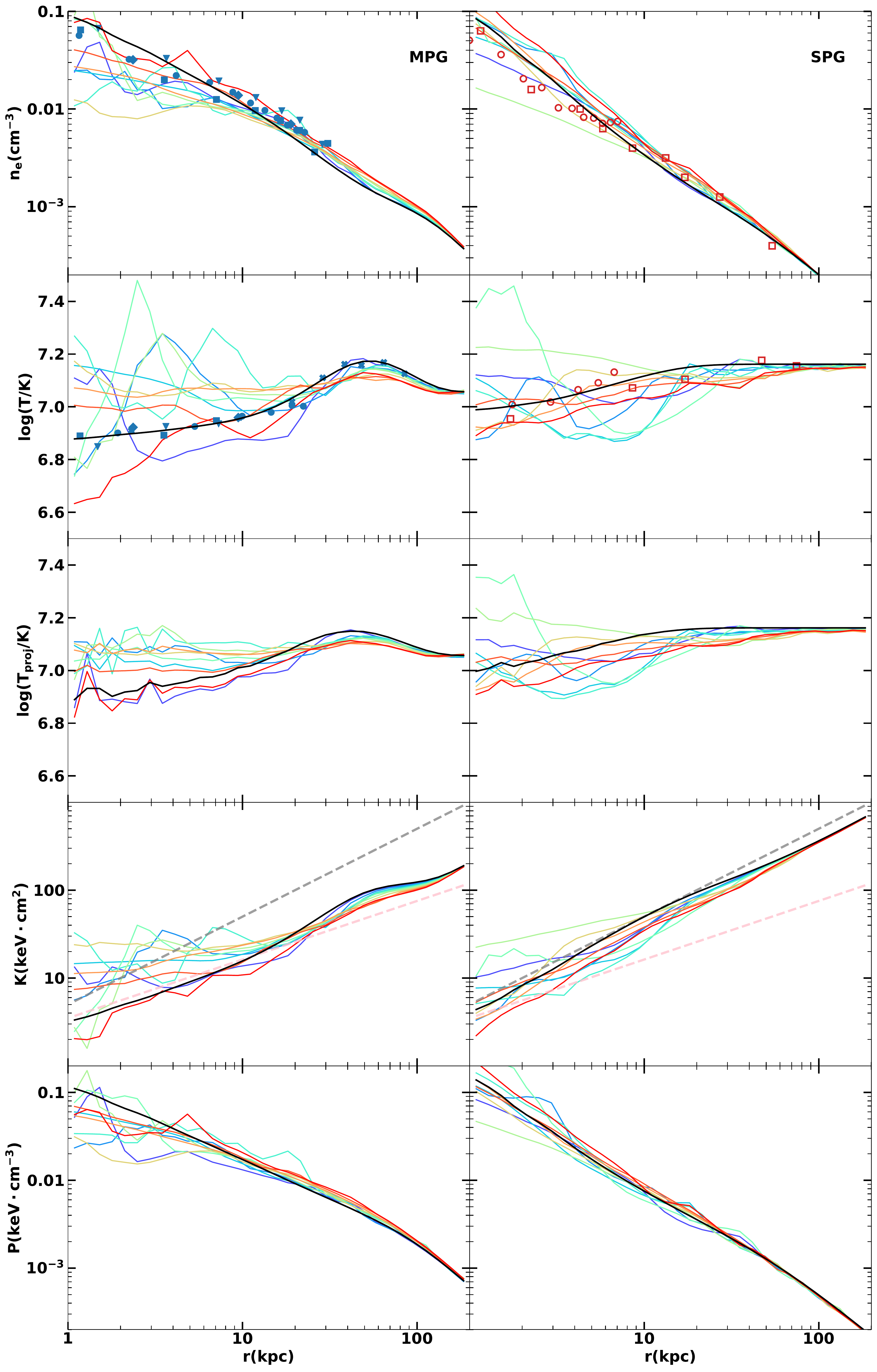} 
       \caption[]{From top to the bottom: the electron density, temperature, projected temperature, entropy, and pressure profiles of the hot gas of the \blue\ (left) and \red\ (right). The density and temperature profiles are weighted by 0.5-9.9 keV X-ray emissivity; the entropy profiles and density profiles are derived from the temperature and density profiles. For each panel, profiles are plotted every 150 Myr, going from red to blue. The black lines show the initial conditions. The data points over-plotted on the density and temperature profiles correspond to the same observational data that is used to generate the initial conditions shown in Figure~\ref{fig:init}.}
     \label{fig:pfs}
  \end{center}
\end{figure*}

Despite the drastic difference between the two galaxies in the spatial extent of the multi-phase gas and the duration of individual feedback episodes, the periodicity appears universal. Both galaxies evolve through cycles of gas cooling and AGN outbursts. Within 1.5 Gyr, the averaged gas properties bounce around the initial condition for both galaxies (Figure~\ref{fig:pfs}). The density, temperature, entropy and pressure profiles of both hot halos stay rather close to their initial conditions. Even though large fluctuations are seen in the ``true'' profiles, the ``observed'' profiles are much smoother due to the projection effect. As an example, we show in the third row of Figure~\ref{fig:pfs} the projected temperature profiles, close to what an observer would see. The projected temperature profiles are smoother and closer to the initial conditions than the profiles directly computed from the simulation data (the second row).

\begin{figure}
  \begin{center}
    \leavevmode
     \includegraphics[width=\columnwidth]{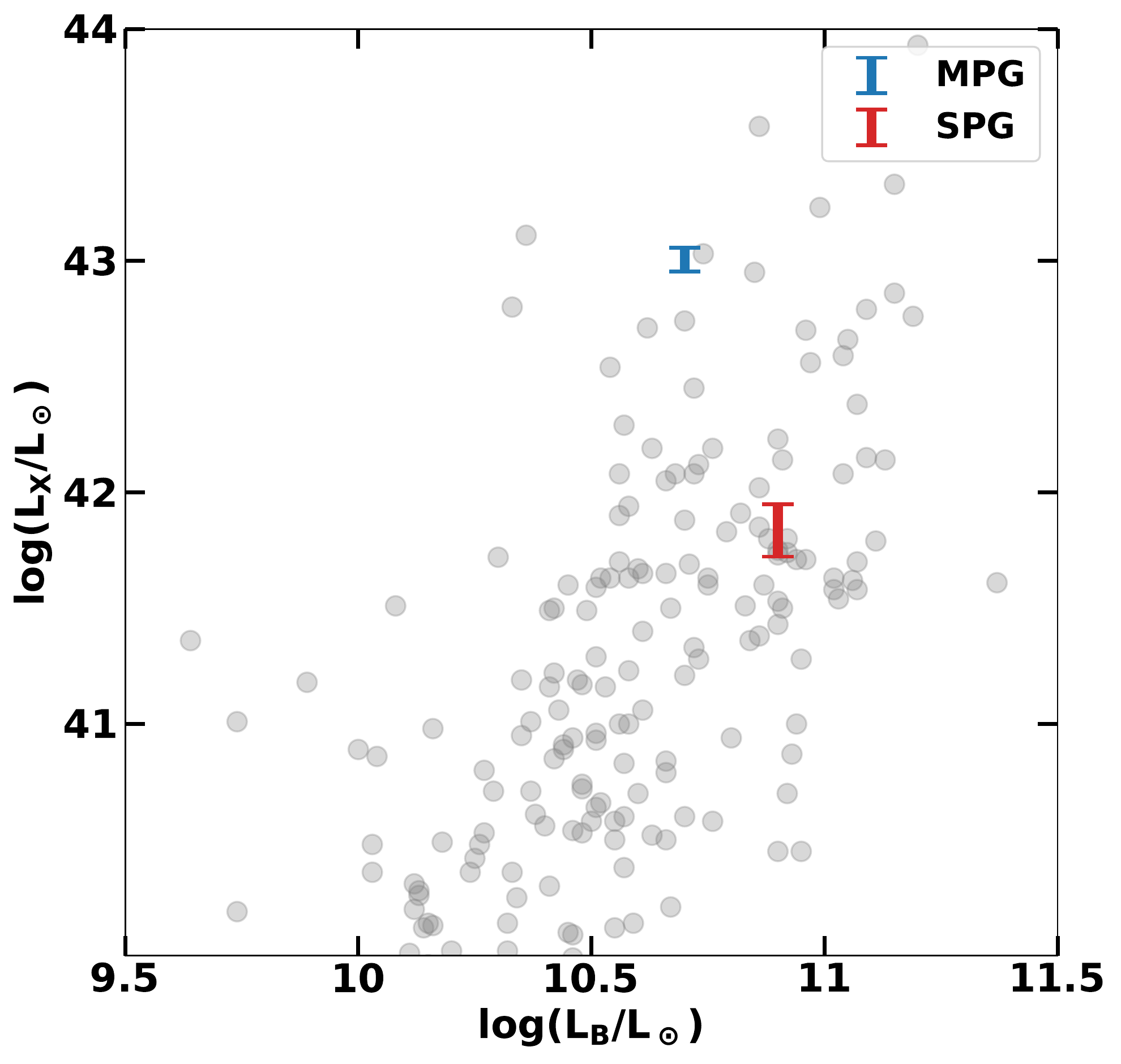} 
       \caption[]{The $L_X-L_B$ relation of elliptical galaxies. The filled gray circles are from \cite{2001MNRAS.328..461O}. The simulated \blue\ and \red\  are shown as the blue and the red symbols, respectively. The error bars represent the range of total X-ray luminosities within 1.5 Gyr.}
     \label{fig:lxlb}
  \end{center}
\end{figure}

The X-ray luminosities of the two systems do not show large fluctuations either. Figure~\ref{fig:lxlb} shows the two galaxies on the $L_X - L_B$ relation. Within 1.5 Gyr of the simulations, the variation of $L_X$ is within 28 percent of the initial condition for the MPG and 49 percent for the SPG. 

\subsection{Thermal Instabilities and multi-phase ISM}\label{ss:result2}

In this section, we analyze our simulation results in more detail, with a focus on the multi-phase ISM in the two galaxies. In particular, we discuss why thermal instabilities develop differently in the two galaxies, and compare our results with analytical work and previous numerical simulations.

The development of thermal instability in a gaseous halo is closely related to its $\min(\tctff)$ ratio. Recent observations suggest that there is a critical ratio of $\min(\tctff)\sim$10 where hot halos of elliptical galaxies and galaxy clusters can develop extended multi-phase gas \citep{Werner2014,2015ApJ...799L...1V}. Numerical simulations also generally agree that thermal instabilities occur when the $\min(\tctff)$ of the system drops below a critical value, typically between a few and 20 \citep{2012MNRAS.420.3174S,2012MNRAS.424..190G,2014ApJ...789...54L,2015ApJ...808...43M}.  

\begin{figure*}
  \begin{center}
    \leavevmode
        \includegraphics[width=.9\textwidth]{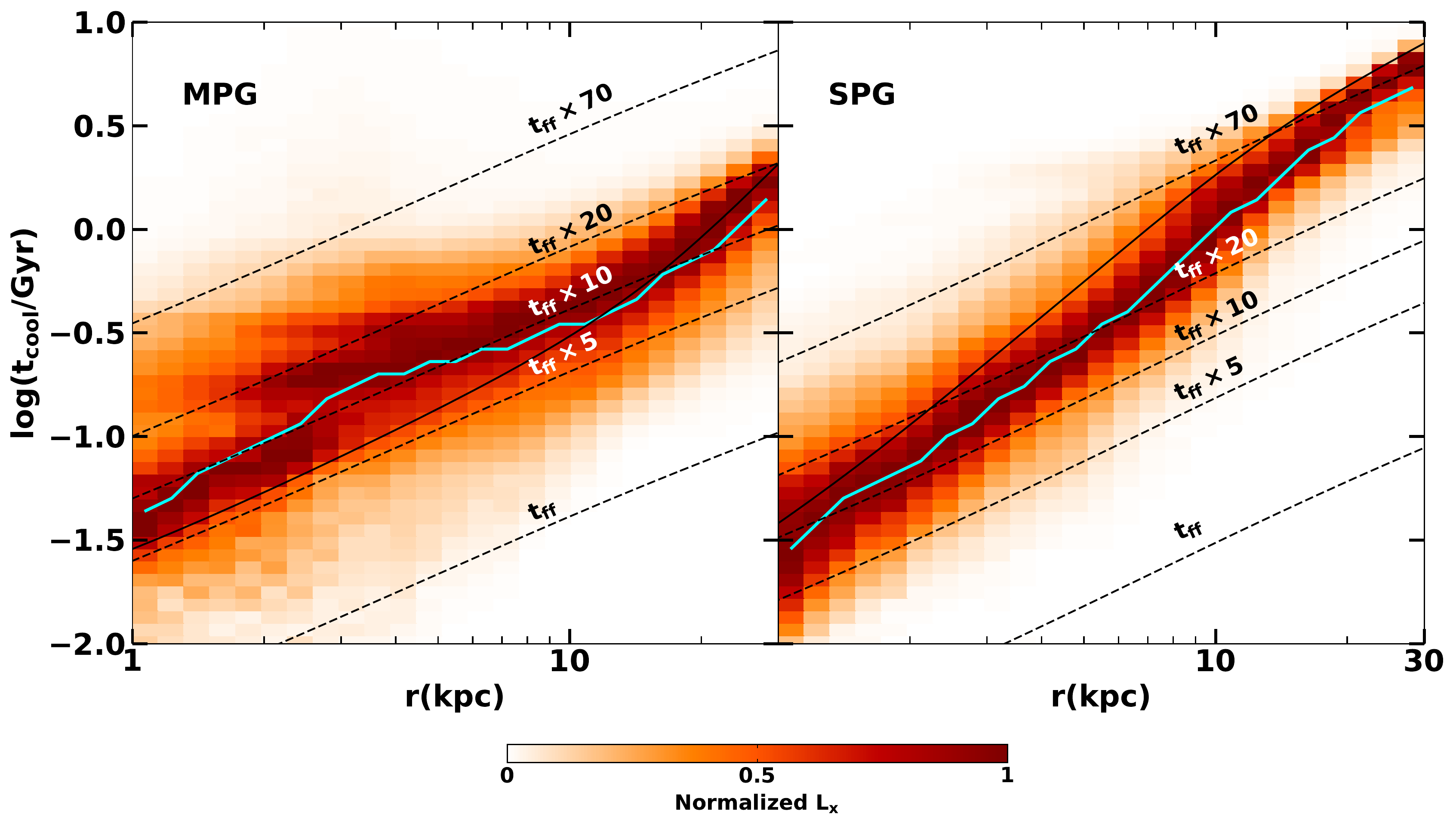} 
       \caption[]{The average radial distributions of the cooling time of hot gas over the first 1.5 Gyr simulation of the \blue\ (left) and the \red\ (right). Plots are made by stacking the 2-D distributions of $t_\mathrm{cool}(r)$ from every simulation output. Color represents the normalized X-ray luminosity within each radial bin. The dashed black dashed lines denote 1, 5, 10, 20 and 70 times of the free-fall time. The solid black line represents the initial $t_\mathrm{cool}(r)$. The cyan line shows the X-ray luminosity-weighted average $t_\mathrm{cool}(r)$. }
     \label{fig:tt_hist}
  \end{center}
\end{figure*}

Our simulations confirm that elliptical galaxies develop extended multi-phase gas when the average $\tctff \approx 10$.  Figure~\ref{fig:tt_hist} shows the ``stacked-average'' $\tctff$ in our two simulated galaxies. For every simulation output (every 10 Myr), we plot the distribution of X-ray luminosity-weighted $t_\mathrm{cool}$ as a function of radius, normalized within each radius bin. We then stack these individual plots to obtain an average distribution. The goal is to create a stacked $t_\mathrm{cool}$ profile similar to that observed in a sample of multi-phase and single-phase ellipticals. We over-plot dashed lines representing 5, 10, 20, and 70 times the free-fall time $t_\mathrm{ff}(r)$. 

In the \blue, the average cooling time (cyan line) of the gas is very close to the $t_\mathrm{ff}\times10$ line (left panel of Figure~\ref{fig:tt_hist}). However, at any radius, the distribution of $t_\mathrm{cool}$ extends well below $5t_\mathrm{ff}$, and the gas that precipitates in the simulation is the gas at the low end of the $t_\mathrm{cool}$ distribution, with a $\tctff$ closer to 1. The same point has also been made in \citet{li2014, 2015ApJ...808...43M} in the context of precipitation in galaxy clusters. 

We have analyzed the physical reason for precipitation occurring when the average $\tctff$ is above 1 in galaxy clusters \citep{li2014, 2014ApJ...789...54L, Voit2017}. The physical processes are similar in our simulations of multi-phase elliptical galaxies discussed here, and therefore we only emphasize the key points without repeating the same analysis: in systems where AGN feedback is the main source of perturbation, AGN itself is the reason for precipitation (sometimes phrased as ``negative feedback''). Both $t_\mathrm{cool}$ and $t_\mathrm{ff}$ increase as a function of radius. Lower entropy gas is uplifted by AGN jets from small radii to larger altitudes where $t_\mathrm{cool}$ ($\sim$ constant if the uplifting process is adiabatic; see \citet{li2014}) becomes comparable to the local $t_\mathrm{ff}$. Besides direct uplifting, AGN jets also drive turbulence, which facilitates precipitation  by suppressing buoyancy damping \citep{Voit2018}. In addition, turbulence enhances density contrast and broadens the distribution of $t_\mathrm{cool}$ at any given radius. The high density, low entropy gas is more likely to cool into cold clouds. 

Based on the {\it Hubble} observations of 77 early-type galaxies, \citet{Lauer2005} hypothesize that dusty clouds form, settle to the center and disappear repeatedly. This cyclical behavior is recreated in our simulated MPG.

In the SPG, however, even the non-linear perturbation from AGN cannot cause precipitation. The average cooling time of the gas is a steep function of radius (right panel of Figure~\ref{fig:tt_hist}). The average $t_\mathrm{cool}$ is above $10t_\mathrm{ff}$ throughout the halo, except at the very center, where the gas does condense. Even the low entropy tail sits mostly above $5t_\mathrm{ff}$. Thus the hot halo of the SPG is too stable for condensation to happen except for the very center.

The stability of hot halo gas is also tightly linked to its entropy defined as $K=kTn_e^{-2/3}$. This is not a surprise as the entropy is almost linearly proportional to cooling time for temperatures we are interested in ($\sim$ 1 keV). The third row of Figure~\ref{fig:pfs} shows the time evolution of the gas entropy profiles in our simulated galaxies. \citet{2015ApJ...803L..21V} analyze a sample of single and multi-phase galaxies and find that the two populations have distinctively different entropy profiles. The single-phase galaxies in their sample follow an outflow solution, with $K(r) = 5r_\mathrm{kpc}~ \mathrm{keV\cdot cm^2}$, shown as the gray dashed line. The entropy of multi-phase galaxies follows a shallower slope, with $K(r) = 3.5r_\mathrm{kpc}^{2/3}~\mathrm{keV\cdot cm^2}$, corresponding to a precipitation threshold at ${\tctff}\approx 10$, shown as the pink dashed line. Our simulations show that the single phase galaxy indeed follows the outflow solution of \citet{2015ApJ...803L..21V} and that the multi-phase galaxy evolves around the precipitation limit. This means that self-regulated AGN feedback is able to maintain the ``single-phase'' or ``multi-phase'' nature of the gaseous halo, and the main reason is that AGN feedback helps maintain the halo in rough hydrostatic equilibrium, and does not cause the galaxy to deviate much from the initial condition.

It is worth emphasizing that AGN feedback does not turn a MPG into a SPG. Even though our simulated MPG experiences brief single-phase moments between feedback cycles, its profile is consistent with that of a typical MPG. Real MPGs likely host extended multi-phase gas even more frequently than our simulated MPG as we do not consider any minor merger events, which could also trigger precipitation. In addition, winds from AGB stars may not fully mix with the hot ISM before they seed further condensation \citep{Joel2008}. 

Whether an elliptical galaxy is single or multi-phase ultimately is determined by its formation history and environment. What our simulations imply is that once an elliptical galaxy has formed, it is locked to its state: a SPG will almost always have a single-phase halo, and a MPG will frequently have extended multi-phase gas and low-level star formation.

\subsection{Cooling, Heating, and Sweeping}\label{ss:result3}

In this section, we examine the balance between cooling, heating, and sweeping of stellar ejecta. Old stellar population in massive elliptical galaxies keeps losing mass into ambient ISM via stellar winds. To prevent the classical cooling flow and excessive star formation, the stellar wind material has to be removed from the galactic interior. This requires energy in addition to the energy needed to account for radiative cooling loss.

Sources of energy injection include AGN feedback, SNIa and thermalized stellar wind itself. The last two terms are modeled as injection of thermal energy in our simulations as described in Section~\ref{sec:methodology}. With the parameters we adopted, heating from SNIa is about 5 times the heating from stellar wind. Our AGN feedback injects mostly kinetic energy at the jet base, which then dissipates into heat via shock waves and turbulence with the former being the dominant channel \citep{Yang2016, Li2017}. The minimum amount of energy needed to sweep stellar ejecta out to radius $r$ is the sum of the change in the potential energy of the stellar ejecta and the enthalpy change.

\begin{figure*}
  \begin{center}
    \leavevmode
      \includegraphics[width=.8\textwidth]{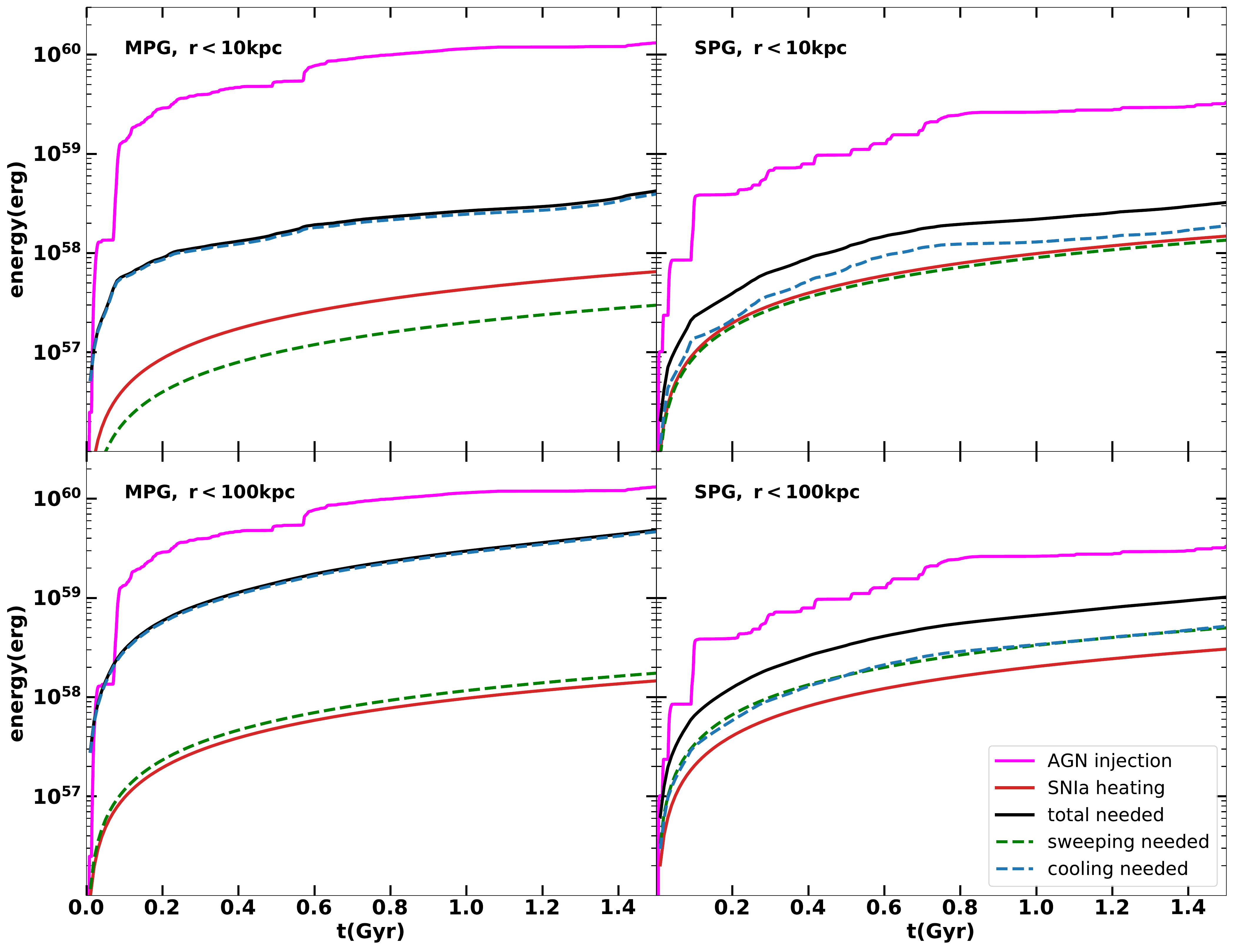} 
       \caption[]{The integrated amount of injected energy in the simulations and the energy needed to compensate for radiative cooling and to sweep the stellar ejecta out to maintain a steady state halo within $r<10~\mathrm{kpc}$ (upper panels) and $r<100~\mathrm{kpc}$ of the simulated \blue\ (left panels) and \red\ (right panels). The sources of energy injection include AGN feedback (magenta lines) and SNIa feedback (red lines). Sources of energy loss including radiative cooling (blue lines) and stellar ejecta sweeping (green lines) are both computed using the initial conditions. }
      \label{fig:budget}
  \end{center}
\end{figure*}

\begin{figure*}
  \begin{center}
    \leavevmode
     \includegraphics[width=0.8\textwidth]{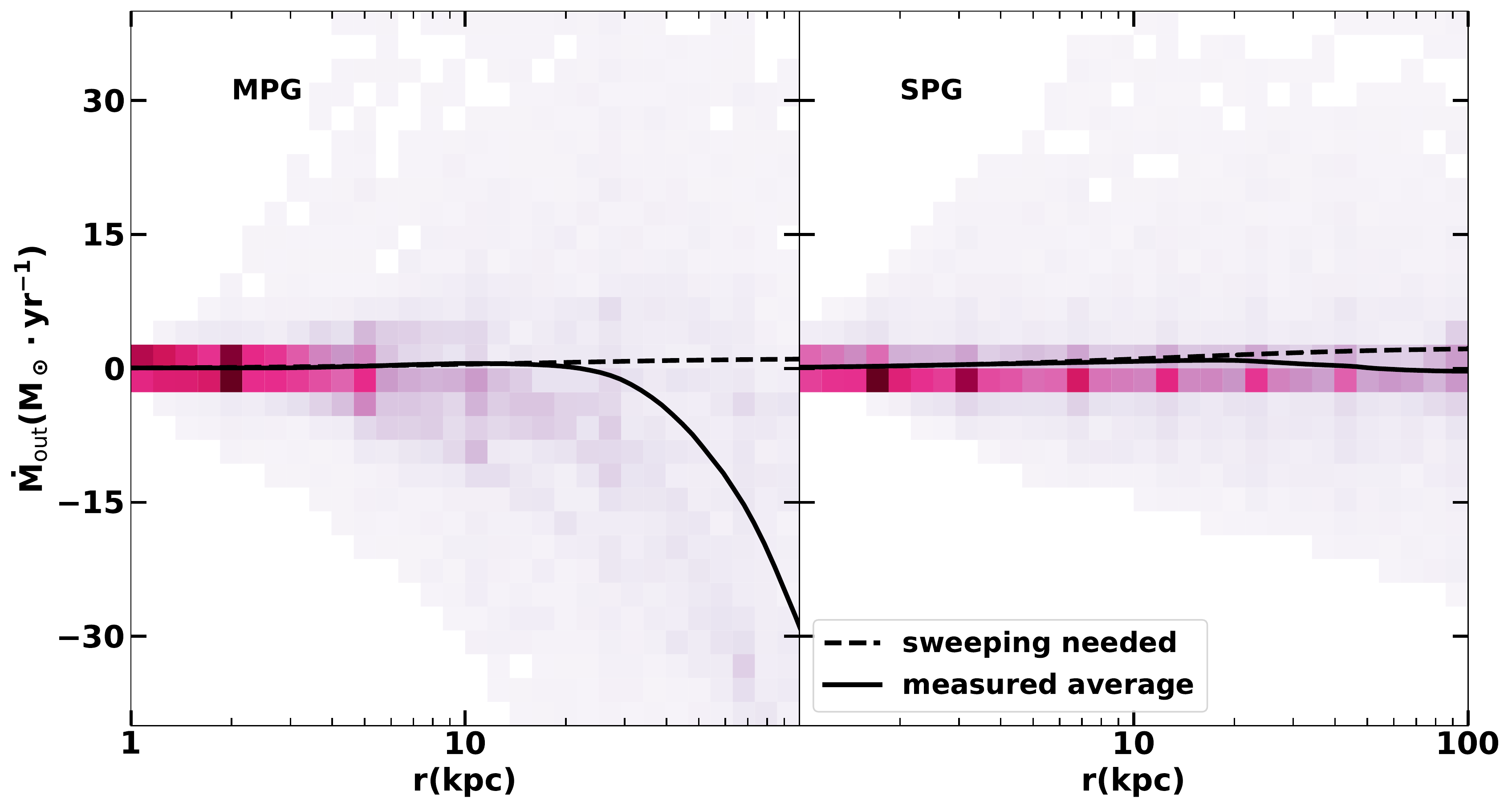} 
       \caption[]{The spherically averaged outflow rates measured from our simulated MPG (left panel) and SPG (right panel). Solid lines show the time-averaged outflow rates and dashed lines show the rates needed to sweep out all the gas produced by old stars to maintain a steady state halo.}
     \label{fig:mdot}
  \end{center}
\end{figure*}

Figure~\ref{fig:budget} shows how different sources of energy loss and injection compare with each other. In the MPG, within 10 kpc (upper left panel), the total energy needed to balance cooling and sweep out stellar ejecta is about 5 times the energy provided by stellar heating. In the SPG, within 10 kpc stellar heating, cooling and sweeping energy requirements are comparable. This is generally consistent with the estimation in \citet{2015ApJ...803L..21V}, which leads to their conclusion that SN-driven outflow effectively sweeps stellar ejecta out of the SPG but not the MPG. When we consider the energy injected by AGN, however, we see a more nuanced picture. In both MPG and SPG, the energy injected by AGN is more than an order of magnitude higher than stellar heating. Although this energy is not all deposited in the central 10 kpc, the work done by AGN has to be comparable to stellar heating in the SPG to maintain the balance within 10 kpc. In the MPG, AGN does most of the sweeping. As pointed out in \citet{2015ApJ...803L..21V}, sweeping in MPG is not effective. Indeed, the total amount of gas that cools ($\sim 5.5\times10^{8}~\mathrm{M_\odot}$) is not much smaller than the amount added by old stars ($\sim 6.9\times10^{8}~\mathrm{M_\odot}$). In other words, only a fraction of the stellar ejecta is swept out; the rest is ``recycled'' and forms multi-phase gas. The fate of the cold gas is discussed in more detail in Section~\ref{sec:discussion3}.

If we consider a larger region of the galaxy ($r<100$ kpc, bottom panels of Figure~\ref{fig:budget}) instead of just the central 10 kpc, we find an even larger deficit if stellar feedback is the only source of heating in both MPG and SPG. Due to its higher gas density, the MPG loses more energy via radiative cooling, which causes the AGN to inject more energy to maintain the balance. Because coupling is not perfect, the total energy injected by the AGN is 2-3 times the energy needed within 100 kpc. This is very similar to what is found in our previous simulations of cool-core galaxy clusters \citep{Li15}. We compare our simulations with cluster simulations in more detail in Section~\ref{sec:discussion3}.

Figure~\ref{fig:mdot} shows the outflow rate that is needed to maintain a steady state halo (dashed line) along with the actual mass flux measured in the simulation. Although the measured flux varies with time, in both MPG and SPG, the average flux matches the expectation almost perfectly within 10 kpc. This is in line with Figure~\ref{fig:pfs} which shows that the properties of the gas in the innermost 10-20 kpc do not deviate much from the initial conditions. However, at $r>10-20$ kpc, the measured outflow rate falls below expectation in both MPG and SPG. As is shown in \citet{Li2017}, the energy deposition of mechanical AGN jets is a steep function of radius ($\sim r^{-3}$) (see also \citet{VoitDonahue2005, Fabian2005}). As a result, swept-up stellar ejecta pile up around $\sim 30$ kpc, gradually increasing the density of the gas (Figure~\ref{fig:pfs}). This then causes the cooling rate to increase, driving an inflow from even larger radii in a process similar to the classical cooling flow. This is why the radial mass flux becomes negative in Figure~\ref{fig:mdot}. Because the MPG has a higher density halo and a shorter cooling time, this effect is more dramatic. We discuss the result of this pileup problem in Section~\ref{sec:discussion1}.

\begin{figure}
  \begin{center}
    \leavevmode
     \includegraphics[width=\columnwidth]{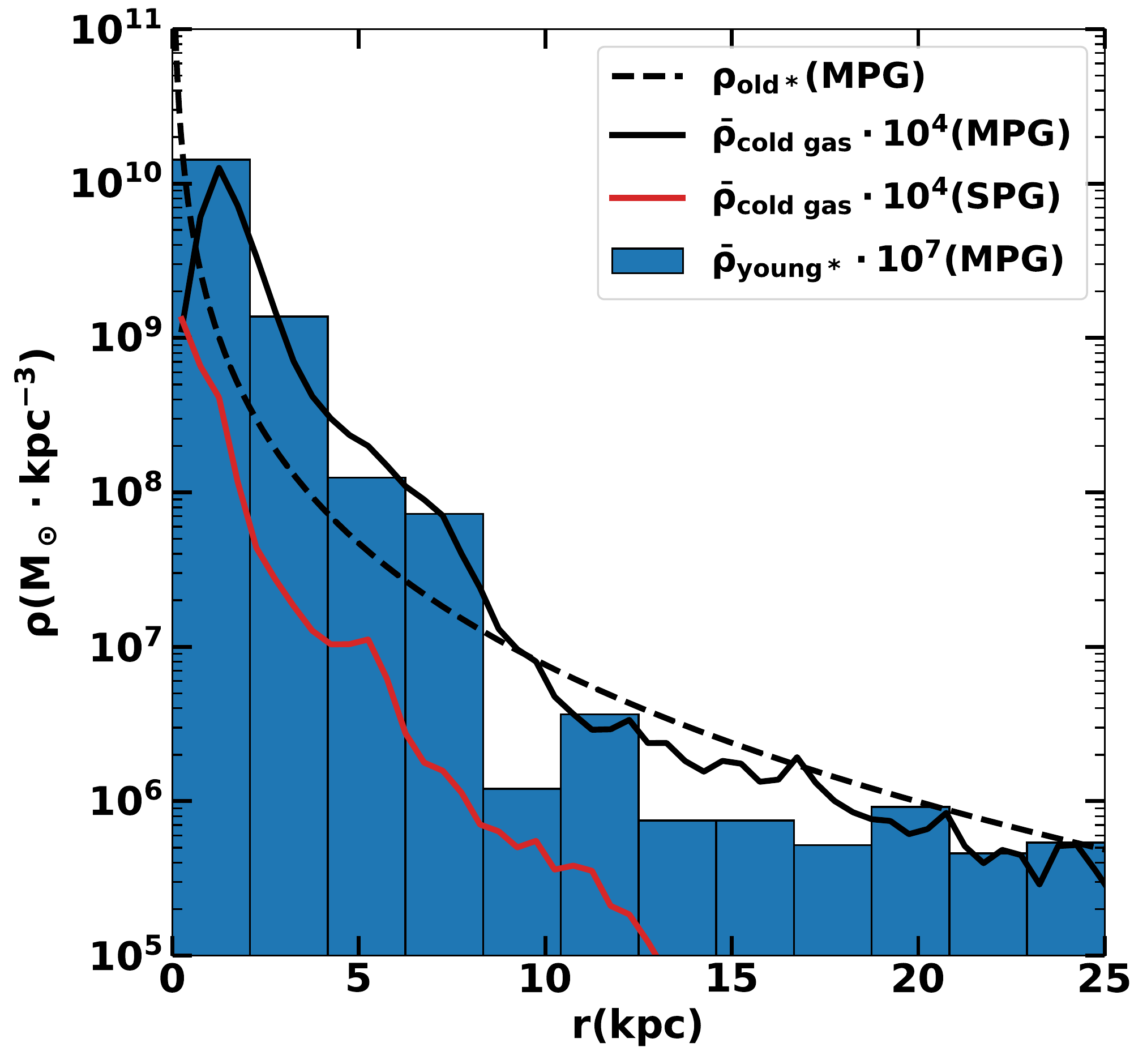} 
       \caption[]{The average density distribution of the cold gas (solid black line), old stars (dashed black line), and young stars (blue histogram) of the \blue. The red line shows the average density distribution of the cold gas in the \red. For presentation purposes, cold gas density and the young stars density are multiplied by $10^7$ and $10^4$, respectively. }
     \label{fig:young_star}
  \end{center}
\end{figure}

\subsection{Young Stars in Multi-phase Elliptical Galaxies}\label{ss:result4}

Low-level star formation in elliptical galaxies has been suggested by observations of optical lines and dust lanes \citep{Lauer2005}. Using ultraviolet Hubble Space Telescope Wide Field Camera 3 imaging, \citet{2013ApJ...770..137F} detected young stars and star clusters in all of the four nearby elliptical galaxies they observed. 

As described in Section~\ref{ss:result1}, stars form in extended multi-phase structures in our simulated MPG, with a rate of $\lesssim 0.1~\mathrm{M_\odot \cdot yr^{-1}}$. We show the radial distribution of cold gas (solid line), old stars (dashed line), and newly formed young stars (histogram) in Figure~\ref{fig:young_star}. The density distributions of cold gas and young stars are again a ``snapshot average'' obtained by stacking the distribution from all simulation output data (generated every 10 Myr). Here young stars are defined as star particles with an age younger than 10 Myr. The old stellar population follows the de Vaucouleurs profile by design (\S \ref{ss:init}). 

To make comparison easier, the young star density is multiplied by a factor of $10^7$ and the cold gas density is multiplied by a factor of $10^4$. The radial distribution of young stars is very similar to the distribution of cold gas. This is not surprising as star formation occurs over a short timescale (the local dynamical time of the cold dense gas). The distribution of young stars is also similar to the old stars, with a slightly steeper slope. 

The star formation rate in our simulated galaxy and more remarkably, the radial distribution of the young stars are in excellent agreement with the measurements in \citet{2013ApJ...770..137F}. 

For comparison, we also plot in Figure~\ref{fig:young_star} the distribution of cold gas in our simulated SPG, which is both much lower in its total amount and much more spatially concentrated than cold gas in the MPG. As mentioned earlier, stars never form in our simulated SPG despite the existence of cold gas. We do caution that our star formation model has been mainly calibrated using simulations of star forming disk galaxies, and thus may not be perfectly suitable for star formation in elliptical galaxies. Nonetheless, even if stars do form in SPGs, they should be very concentrated in the nuclei. Our model predicts very different spatial distributions of both multi-phase gas and stars in single-phase and multi-phase galaxies, which can be tested with future observations of a larger sample of ellipticals.

\subsection{Velocity Dispersion of the Hot Gas}\label{ss:result5}

Besides the obvious difference in the morphology of cold gas between the MPG and SPG, there is also significant difference in the AGN duty cycle and power, and thus the level of perturbation it causes in the hot ISM.

As described in Section~\ref{ss:result1} and shown in Figure~\ref{fig:time_evo}, in the MPG, AGN feedback has a longer duty cycle, longer ``on'' time, and overall more energy output than in the SPG (see also Figure~\ref{fig:budget}). As a result, the MPG also has a more perturbed hot halo. This is reflected in the profiles of the hot gas shown in Figure~\ref{fig:pfs}. The gas density, temperature, entropy, and pressure of the MPG generally show larger fluctuations than in the case of the SPG. 

\begin{figure}
  \begin{center}
    \leavevmode
    \includegraphics[width=\columnwidth]{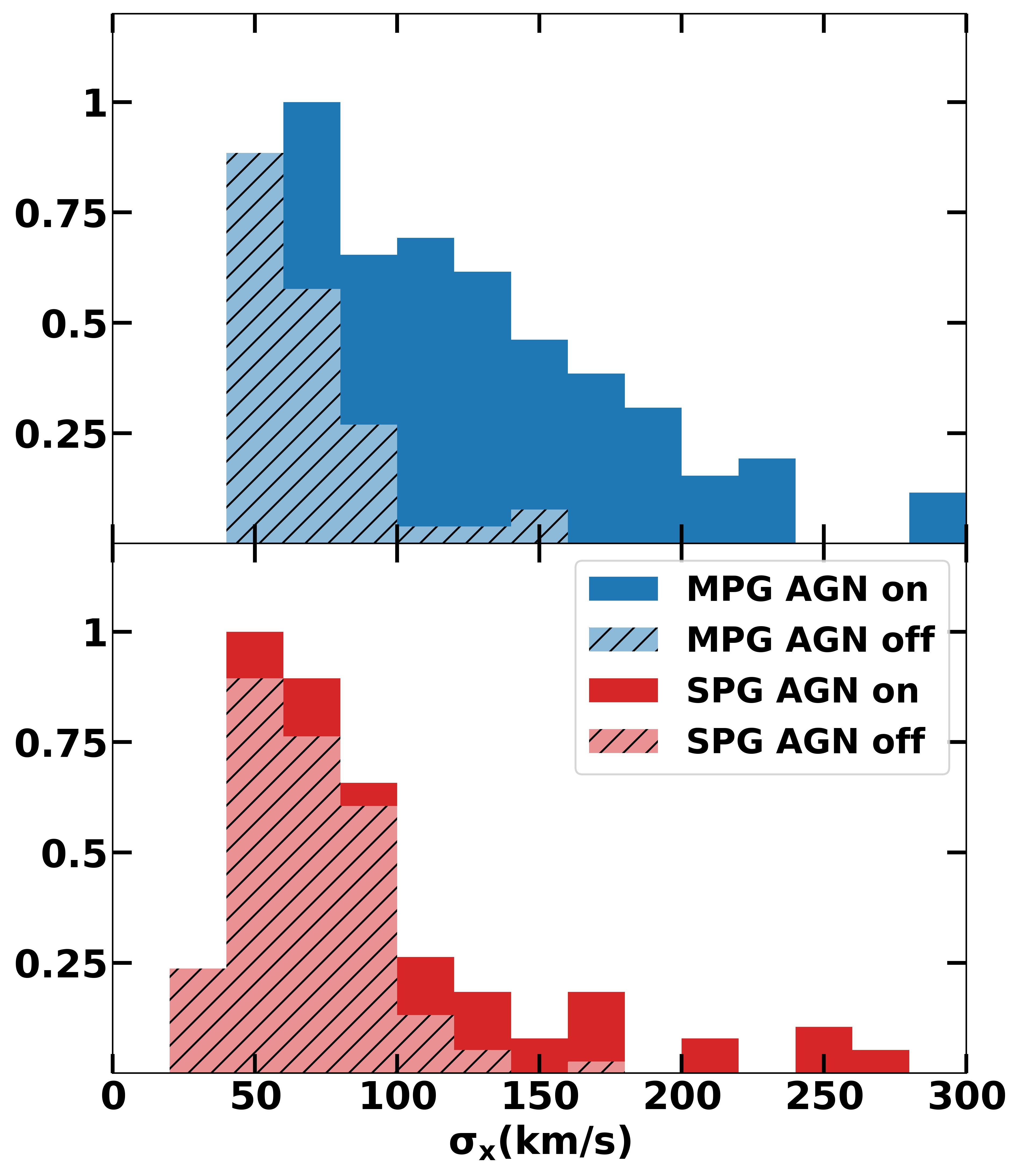}  
       \caption[]{Normalized distributions of velocity dispersion (computed as the standard deviation of velocities along $x-$axis) of $r<10~\mathrm{kpc}$ region weighted by the 0.5-9.9 keV X-ray emissivity in our simulated MPG (top panel) and SPG (bottom panel). Hatched regions denote the epochs when the AGN is off.}
     \label{fig:turb}
  \end{center}
\end{figure}

Figure~\ref{fig:turb} shows the distribution of average line-of-sight velocity dispersion of the central 10 kpc region of our simulated MPG and SPG. Though the range of $\sigma$ is similar ($\sim 50 - 300~\mathrm{km\cdot s^{-1}}$) for both galaxies, MPG show a higher fraction time with $\sigma > 100~\mathrm{km\cdot s^{-1}}$. We further define an active phase by selecting times when the AGN is on. Both galaxies show higher sigma during their active times as one would expect, and similar $\sigma$ distribution when the AGN is off, typically between 50 and 100 $\mathrm{km\cdot s^{-1}}$. Selecting times when cold gas is present within $r <10$ kpc as the active phase yields similar results. These findings are generally consistent with previous 3D and 2D simulations of similar systems under the influence of momentum-driven AGN feedback \citep{2012MNRAS.424..190G, 2015MNRAS.448.1979V}. 

\citet{Irina2017} measured turbulent velocities of 13 nearby elliptical galaxies using resonance scattering and line broadening. Their measured range and mean ($\sim 110~\rm km\cdot s^{-1}$) are in good agreement with our simulations. In particular, they obtained a $\sigma_\mathrm{1D}=172_{-79}^{+108}~\mathrm{km\cdot s^{-1}}$ for NGC 5044 (our MPG); and even though additional data is needed for better measurements of turbulence in NGC 4472 (our SPG), their analysis supports weak turbulence, in good agreement with our simulation results. With a larger sample of elliptical galaxies, we should be able to test our prediction that velocity dispersion of the hot halo gas is positively correlated with AGN activities and the existence of cold gas.

\section{Discussion}\label{sec:discussion}
\subsection{Long-term Evolution}\label{sec:discussion1}

In this section, we discuss the evolution of the galaxies after the first 1.5 Gyr. As discussed in Section~\ref{ss:result3}, because AGN-driven galactic wind weakens at larger radii, stellar ejecta gradually pile up around a few tens of kpc, which then increases the cooling rate. As a result, a major precipitation event happens at $t\sim 1.4$ Gyr in the MPG, with $\sim 2\times10^{9}~\mathrm{M_\odot}$ of cold gas cooling out of the hot halo. The cold gas falls to the center and forms a massive cold disk. The inner boundary of the disk extends to the accretion region. The disk keeps feeding the central black hole, continuously powering AGN feedback, but the angular momentum prevents it from being quickly accreted. The hot ISM cools directly onto the disk. The central few kpc region gets over-heated and becomes isentropic. Eventually, after about 2 Gyr, star formation and SMBH accretion completely consume the cold disk and the system returns to a state similar to its initial condition. The SPG experiences a similar cycle, at a much later time ($t \sim 4\,\mathrm{Gyr}$). 

One can argue that AGN feedback is still maintaining a thermal balance during the 2 Gyr cold disk phase. However, the existence of an overly massive cold disk and a large isentropic core violate the observations of most elliptical galaxies. Thus, we consider this phase unphysical and do not include it when discussion our main results in Section~\ref{sec:results}. A persisting massive cold disk is commonly seen in numerical simulations of massive galaxies or clusters \citep{2014ApJ...789...54L, Prasad2015, 2017MNRAS.468..751E}. Below we discuss the possible solutions to this problem.

The solutions to the cold disk phase fall in three categories: (i) a major cooling event should never happen, 
or (ii) the disk should never form in such an event, or (iii) the cold disk is a short-lived phase. It is possible that minor mergers, which are not included in our idealized simulations, can provide heating via dynamical friction in the outskirts of the galaxies at $r\sim 30$ kpc to keep the gas properties constant. It is also possible that our AGN feedback is too simplistic, and real AGN jets are able to more effectively remove gas accumulated due to stellar winds to large radii out of the galaxy.

More probably though, elliptical galaxies do periodically host a significant amount of cold gas and go through a very short quasar phase \citep{2017arXiv171204964Y}. It is possible that the newly formed cold clouds are more easily destroyed due to thermal conduction or they have additional pressure support from cosmic rays and/or magnetic fields (thus a larger ram-pressure cross-section for given mass) \citep{Li2018}, and therefore a massive disk never forms. It is also possible that massive cold disks do form \citep{Alatalo2013} and trigger a quasar as a result of the secular evolution of the galaxies even without a wet merger. Quasar-mode AGN feedback could effectively destroy or remove the cold gas via radiation-driven dusty winds \citep{2017ApJ...835...15C}. Cosmic rays from supernovae could also drive galactic winds and remove gas from such disks \citep{2013ApJ...777L..16B,SalemBryan14,2016ApJ...824L..30P,2017ApJ...834..208R,2018arXiv180306345B}. Cosmic rays injected by the AGN could disperse in the circumgalactic medium and in the process heat the gas via the streaming instability \citep{2017ApJ...844...13R}. Alternatively, we simply do not have the resolution to resolve the instabilities of the cold disk for it to be quickly accreted. It is also possible that the longevity of the cold disk has to do with numerical over cooling at the mixing layers between the hot gas and cold gas \citep{Brighenti2015}. 

Regardless of the true reason for the existence of the disks, we think that the cold disk should be short-lived if it ever forms in real galaxies as elliptical galaxies are rarely observed to host a cold disk of $10^9~\mathrm{M_\odot}$, and the system should quickly return to the cycles described in Section~\ref{sec:results}. We will test some of the hypothesizes above to solve the disk problem in future work.

\subsection{Resolution and Model Parameters}\label{sec:discussion2} 

\subsubsection{Lower Resolution Run}\label{sec:discussion2a}

To test the convergence of our simulations, we perform lower resolution runs with a maximum refinement level of 10, one level coarser than our standard runs. Everything else is kept the same as in the standard run, including the physical size of the jet launching plane and the accretion zone. We find that the results are qualitatively very similar to our standard runs. In particular, we find similar cyclical behavior (although the exact durations differ) and $\tctff$ distribution. The main difference is that there are fewer cold clumps and individual clumps are larger, as one would expect. 

\subsubsection{Testing Model Parameters}\label{sec:discussion2b} 

Since our simulation results are converged, all our tests of the impact of changing model parameters are performed using the lower resolution with a maximum refinement level of 10. We find that our simulation results do not change qualitatively when we change the kinetic fraction of the AGN feedback energy as long as the feedback is not purely thermal. This is consistent with what is found in cluster simulations in \citet{2014ApJ...789...54L, 2017ApJ...841..133M}. The result does not depend on the exact value of small angle precession either, again in agreement with \citet{2014ApJ...789...54L, 2017ApJ...841..133M}. However we do find that the cold disk forms earlier when there is no precession at all. Note that even though the general results are robust, the details of the galaxy evolution are very sensitive to any small change in simulation parameters or initial conditions. These details include the exact duration of individual cycles, the exact amount of cold gas that cools in each cycle, and the exact morphology of the extended multi-phase structures, which can be more bi-polar, isotropic, or more disky. This is because the evolution of the system is highly non-linear. Thus, the results discussed earlier, based on our standard run, should not be seen as the one and only evolutionary path, but rather as a typical one.

The MPG simulation without star formation leads to similar results to the standard run, because star formation is inefficient and has little impact on the evolution of the system. This is quite different from our previous simulations of cool-core clusters and is discussed in more detail in Section~\ref{sec:discussion3}.

The one parameter that can alter the results significantly is the feedback efficiency $\epsilon$. When we increase $\epsilon$ by a factor of 2 to $1\%$, we see very little change compared with our standard run, but when we decrease $\epsilon$ to $0.1\%$, the simulated galaxies appear drastically different. A large amount of gas ($\sim10^{8}~\mathrm{M_\odot}$) cools into spatially extended structures and forms stars even in the SPG. This is not surprising as the total energy injected by the SMBH in our standard run is only 2-3 times the required energy to balance radiative cooling and sweep stellar ejecta out of the halo (Figure~\ref{fig:budget}). A lower $\epsilon$ causes the galaxies to evolve significantly away from their initial conditions and thus violate the observations. This finding is in tension with \citet{2012MNRAS.424..190G}, which we discuss in detail in Section~\ref{sec:discussion4}.

\begin{figure}
  \begin{center}
    \leavevmode
    \includegraphics[width=\columnwidth]{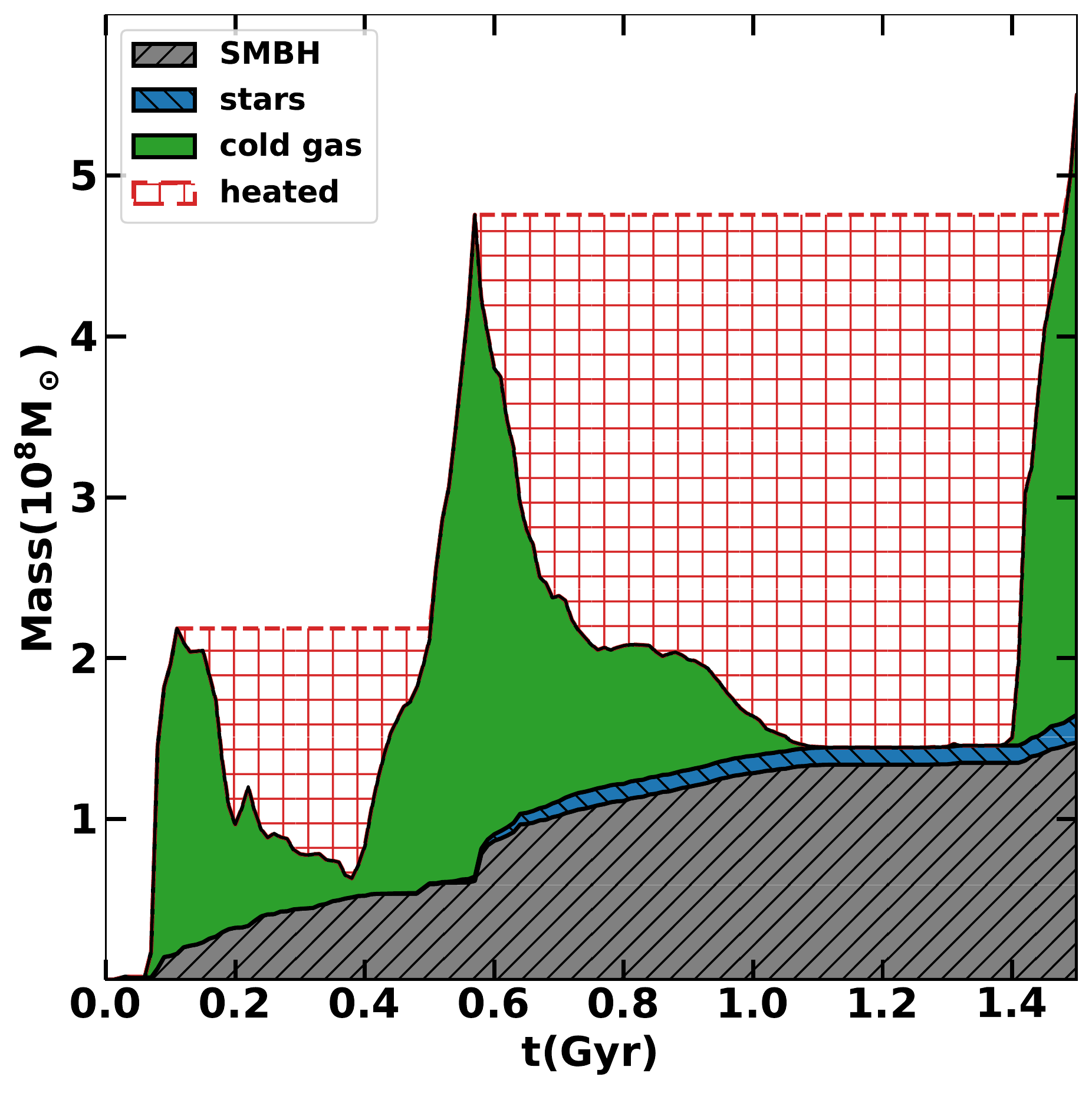}  
       \caption[]{The integrated amount of cold gas ($T<10^5$ K) that is processed by the SMBH (grey), the gas that has turned into young stars (blue), the cold gas that exists in the halo of the simulated MPG (green), and the lower limit of the cold gas that is returned to the hot phase via shredding and mixing (red).}
     \label{fig:coldmass}
  \end{center}
\end{figure}

\subsection{Comparison with Cluster Simulations} \label{sec:discussion3}
The evolution of our simulated elliptical galaxies, especially the MPG, shares many similarities with simulated cool-core galaxy clusters in our previous works \citep{2014ApJ...789...54L,li2014,Li15}. Both MPG and cool-core clusters experience cycles of AGN outbursts, formation of extended multi-phase gas and star formation. The criterion for thermal instability to develop in both systems is $\tctff\sim 10$, and the feedback efficiency required for AGN to balance cooling is also similar ($\epsilon \sim 0.5-1\%$). 
The amount of gas that precipitates and the number of stars that form are lower in MPG than in clusters. As a result, individual cycles are also shorter in MPG than in clusters. This is simply because the density of the hot gas in the MPG is much lower than in the center of a cool-core clusters. Thus, the amount of low entropy hot gas that can cool in each cycle is smaller. 

The most striking difference between the MPG and the cluster simulations is the fate of the cold gas. Figure~\ref{fig:coldmass} shows the total amount of cold gas ($T<10^5$ K) processed by the SMBH, the amount that formed stars, and the gas that exists in the system integrated over time. The amount of gas turning into stars is only $\sim 1/10$ of that falling onto the SMBH. This is quite the opposite of what is seen in our simulated cool-core cluster, where the amount of star formation is an order of magnitude larger than SMBH accretion. 

We speculate that this is because the cold clumps in our simulated MPG are ``fluffier'' (with lower average density and pressure) than in clusters, and thus are less effective at turning into stars. They are likely also more susceptible to ram pressure stripping and shredding. It is difficult to quantify the amount of cold gas that is mixed back into the hot phase in a grid-based code. However, we can obtain a lower limit on the amount of heated gas based on the simple fact that the total amount of cold gas formed in the galaxy has to increase monotonically. As Figure~\ref{fig:coldmass} shows, a significant amount of cold gas is returned to the hot phase before it forms stars or falls onto the SMBH, often more than the amount of traceable cold gas.

Our simulations suggest that star formation efficiency is lower in MPGs than in cool-core clusters which has similar efficiency as the main sequence galaxies \citep{Li15}. The average gas depletion time in our simulated MPG is about $7 \,\mathrm{Gyr}$. However, we caution that neither the cold clumps nor star formation is well resolved in our simulations. Our conclusion needs to be confirmed with dedicated future high-resolution studies.

\subsection{Comparison with Other Works}\label{sec:discussion4}
In this section, we compare our simulation results with two other works: \citet{2012MNRAS.424..190G} and \cite{2017MNRAS.468..751E} that study AGN feedback in idealized elliptical galaxies using 3D hydro simulations.

The setup of our MPG is very similar to \citet{2012MNRAS.424..190G} who use the FLASH code. We both find that cold-mode accretion powered mechanical AGN feedback can successfully suppress cooling in elliptical galaxies. The spatial extent of the cold gas and the level of turbulence in the hot halo in our MPG are generally in good agreement with \citet{2012MNRAS.424..190G}. A remarkable difference is that \citet{2012MNRAS.424..190G} favor a much lower SMBH feedback efficiency $\sim 10^{-4}-10^{-3}$. We attribute this difference mainly to the difference in the treatment of cold gas in the two simulations. \citet{2012MNRAS.424..190G} remove cold gas using a dropout term and assumes that all the gas that cools falls onto the SMBH. We follow the evolution of cold clumps in our simulations and find that only a fraction ($<1/5$) of cold gas actually gets processed by the AGN, and a larger fraction is mixed back into the hot phase before it is accreted or forms stars (see Figure~\ref{fig:coldmass} and Section~\ref{sec:discussion3}).

\cite{2017MNRAS.468..751E} uses smoothed particle hydrodynamics code to study the quenching effect of AGN feedback in massive elliptical galaxies. Despite the differences in the simulation codes, initial setup, and modes of AGN feedback (\cite{2017MNRAS.468..751E} includes radiative feedback in addition to mechanical feedback), we find generally consistent results. In particular, both works find that AGN feedback has only a small effect on the $L_X$ of the galaxy, and thus causes little change to $L_X$ position on scaling relations (e.g., the $L_X-L_B$ relation in Figure~\ref{fig:lxlb}). \cite{2017MNRAS.468..751E} also find that AGN feedback is most effective in pushing gas out of the central $\sim 30$ kpc, and a persisting cold disk forms in all of their simulations. This suggests that the formation of the disk at later times in our simulations (Section~\ref{sec:discussion1}) does not depend on the code or the specific implementation of the feedback model (at least in these simulations).

\section{Conclusions}\label{sec:conclusion}

We have performed 3D AMR hydrodynamical simulations of two idealized elliptical galaxies based on the observations of NGC 5044 and NGC 4472. We choose the former as a representative multi-phase elliptical galaxy (MPG, defined as a galaxy hosting spatially extended multi-phase gas) and the latter a representative single-phase galaxy (SPG, defined as a galaxy that has no cooler component or the cooler gas is only detected in the nucleus). We model momentum-driven AGN feedback powered by cold-mode accretion, and study the interplay between radiative cooling, AGN feedback and feedback from old stars in the galaxies. We focus on the development of thermal instabilities in the two galaxies and the formation of multi-phase gas. We compare our simulation results with previous analytical prediction and observations of nearby elliptical galaxies including the velocity dispersion of the hot gas and the distribution of young stars. Our main results are summarized below:
\begin{enumerate}

\item The simulated MPG has an average $\tctff\sim 10$ within $\sim 30$ kpc, and momentum-driven AGN jets frequently trigger thermal instabilities in the halo, causing hot gas to cool into extended multi-phase structures. The galaxy experiences precipitation-regulated AGN feedback cycles with typical periods of a few hundred Myr. In the simulated SPG, $\tctff$ is a steeper function of radius, and is always above 10 except in the very center where cooling can happen. The galaxy rarely shows extended multi-phase gas, and AGN feedback cycles are typically only a few tens of Myr. 

\item Most of the stellar ejecta in the inner halo of the SPG are swept out, while some of them cool in the MPG, As predicted in \citet{2015ApJ...803L..21V}. Most of the sweeping work is done by AGN feedback rather than stellar feedback in both MPG and SPG. Because shock heating is a steep function of radius, AGN-driven outflow does not result in a steady state halo. Instead, it slows down around 30 kpc, causing gas to pile up. As a result, after a few Gyr, a large a mount of gas precipitates and forms a massive cold disk.

\item AGN feedback is on more frequently with a higher average power in the MPG than in the SPG. As a result, the hot halo gas in the MPG shows a high velocity dispersion ($>100~\mathrm{km\cdot s^{-1}}$) more frequently than the SPG. Both galaxies have similar velocity dispersions ($\sim 50-100~\mathrm{km\cdot s^{-1}}$) during the quiescent phase when the AGN is off.

\item The spatially extended multi-phase gas in the simulated MPG is often associated with extended star formation. The star formation rate is low ($< 0.1~\mathrm{M_\odot\cdot yr^{-1}}$). The spatial distribution of the young stars is similar to the cold stellar population, but slightly steeper as a function of radius. This is in excellent agreement with the Hubble observations of young stars in several nearby elliptical galaxies. Star formation does not occur in our simulated SPG.

\item Compared to our previous simulations of cool-core galaxy clusters, we find that in elliptical galaxies, the thermal instability criterion ($\tctff\sim 10$) and the required AGN feedback efficiency ($\epsilon \sim 0.5-1\%$) are very similar. The main difference between the simulated elliptical galaxies and cool-core clusters is the fate of the cold gas. In elliptical galaxies, cold gas has a lower average density and thus forms stars less efficiently. In addition, in multi-phase elliptical galaxies, a larger fraction of cold gas is shredded and returned to the hot phase before it can be accreted onto the SMBH. 
\end{enumerate}
Our simulations suggest that maintenance-mode AGN feedback not only maintains the general quiescent state of massive elliptical galaxies, but also maintains the multi- or single-phase nature of the galaxy. In other words, once an elliptical galaxy is formed with a halo structure that allows local thermal instability to develop (with a flatter entropy or $\tctff$ profile), it will frequently exhibit extended multi-phase gas, whereas a galaxy with a steeper entropy or $\tctff$ slop will only periodically have cold gas in its very center. Future simulations with more physical processes included (e.g., cosmic rays) will help to better understand the long-term evolution of the galaxies.\\ 

\section*{acknowledgements}

MR acknowledges financial support from NASA ATP 12-ATP12-0017 grant and NSF grant AST 1715140. The authors acknowledge the computational resources from XSEDE. 
The simulations were performed and analyzed using the open-source code ENZO and python package yt \citep{yt}, which are the products of international collaborations of many independent scientists from numerous institutions. Their commitment to open science has helped make this work possible.

%%%%%%%%%%%%%%%%%%%%%%%%%%%%%%%%%%%%%%%%%%%%%%%%%%



\bibliographystyle{mnras}
\bibliography{template} % if your bibtex file is called example.bib


%%%%%%%%%%%%%%%%%%%%%%%%%%%%%%%%%%%%%%%%%%%%%%%%%%

%%%%%%%%%%%%%%%%% APPENDICES %%%%%%%%%%%%%%%%%%%%%

%%%%%%%%%%%%%%%%%%%%%%%%%%%%%%%%%%%%%%%%%%%%%%%%%%

% Don't change these lines
\bsp	% typesetting comment
\label{lastpage}
\end{document}